\documentclass[pdflatex,sn-nature]{sn-jnl}

\usepackage{graphicx}
\usepackage{multirow}
\usepackage{amsmath,amssymb,amsfonts}
\usepackage{amsthm}
\usepackage{mathrsfs}
\usepackage[title]{appendix}
\usepackage{xcolor}
\usepackage{textcomp}
\usepackage{manyfoot}
\usepackage{booktabs}
\usepackage{algorithm}
\usepackage{algorithmicx}
\usepackage{algpseudocode}
\usepackage{listings}

\usepackage{siunitx}

\usepackage{hyperref}
\hypersetup{colorlinks, citecolor=black, filecolor=black, linkcolor=black, urlcolor=black}

\pdfoutput=1

\usepackage[utf8]{inputenc}
\usepackage[T1]{fontenc}
\usepackage{mathptmx}
\usepackage{etoolbox}

\begin{document}

\title{Metasurfaces for infrared multi-modal microscopy: phase contrast and bright field}

\author*[1]{Shaban B. Sulejman}\email{sulejmans@unimelb.edu.au}

\author[1]{Lukas Wesemann}

\author[2]{Mikkaela McCormack}

\author[1]{Jiajun Meng}

\author[3]{James A. Hutchison}

\author[1]{Niken Priscilla}

\author[4]{Gawain McColl}

\author[5]{Katrina Read}

\author[6]{Wilson Sim}

\author[7]{Andrey A. Sukhorukov}

\author[1,8]{Kenneth B. Crozier}

\author*[1]{Ann Roberts}\email{ann.roberts@unimelb.edu.au}

\affil[1]{ARC Centre of Excellence for Transformative Meta-Optical Systems, School of Physics, The University of Melbourne, Victoria 3010, Australia}

\affil[2]{Dorevitch Pathology, Heidelberg, Victoria 3084, Australia}

\affil[3]{ARC Centre of Excellence in Exciton Science, School of Chemistry, The University of Melbourne, Victoria 3010, Australia}

\affil[4]{Florey Institute of Neuroscience and The University of Melbourne, Parkville, VIC 3010, Australia}

\affil[5]{St Vincent's Private Hospital, East Melbourne, Victoria 3065, Australia}

\affil[6]{School of Science, RMIT University, Melbourne, Victoria 3001, Australia}

\affil[7]{ARC Centre of Excellence for Transformative Meta-Optical Systems (TMOS), Department of Electronic Materials Engineering, Research School of Physics, The Australian National University, Canberra, ACT 2601, Australia}

\affil[8]{ARC Centre of Excellence for Transformative Meta-Optical Systems, Department of Electrical and Electronic Engineering, The University of Melbourne, Victoria, 3010, Australia}


\abstract{Different imaging modalities are used to extract the diverse information carried in an optical field. Two prominent modalities include bright field and phase contrast microscopy that can visualize the amplitude and phase features of a sample, respectively. However, capturing both of these images on the same camera typically requires interchanging optical components. Metasurfaces are ultra-thin nanostructures that can merge both of these operations into a single miniaturized device. Here, a silicon-based metasurface that supports a Mie resonance is demonstrated to perform near-infrared phase contrast and bright field multi-modal microscopy that can be tuned by changing the polarization of the illumination. We performed experiments using optical fields with phase variations synthesized by a spatial light modulator and introduced by propagation through semi-transparent samples, including \textit{C. elegans}, unstained human prostate cancer cells and breast tissue. The results demonstrate the potential of metasurfaces for label-free point-of-care testing.}

\keywords{Metasurfaces, microscopy, phase contrast imaging, biological imaging, Mie resonances, image processing}

\maketitle
\newpage 



\section*{Introduction}\label{section:introduction}


Multi-modal microscopy is a powerful tool for acquiring images with a diverse range of imaging techniques using a single multi-functional system \cite{Chong2013, Harfouche2023}. This includes, but is not limited to, confocal microscopy \cite{Kuppers2023}, super-resolution imaging \cite{VillegasHernandez2022, Descloux2018}, tomography \cite{Schulz2012} and fluorescence microscopy \cite{Zheng2022}. Its versatility has been evident across a wide range of applications, from analyzing meteorite structures \cite{Lo2019} to advanced biomedical imaging \cite{Monkemoller2015, Zhang2018}. Bright field microscopy stands as the simplest and most widely used imaging technique for examining microscopic samples that introduce amplitude variations through absorption. Despite its popularity, it cannot capture the phase variations of light essential for visualizing transparent samples, such as unstained cells. Phase microscopy techniques, including Zernike \cite{Zernike1942} and differential interference contrast (DIC) microscopy \cite{Lang1982}, can detect features attributed to varying refractive index or thickness in transparent samples and quantitative methods can measure these phase variations \cite{Park2018}. While various methods of phase microscopy have been integrated with other imaging modalities \cite{Zheng2018, Kumar2020, Dubey2018, Zhang2017, Brombal2023}, its integration with bright field microscopy provides a comprehensive sample visualization \cite{Paganin2002, Yang2019, Wu2019, PicazoBueno2023, Byeon2016} and information that can be key to quantitative phase imaging strategies \cite{McReynolds2017}. However, the complexity, size and cost of the associated equipment have largely limited their availability to specialized laboratories, precluding access to point-of-care diagnostics and in developing countries or remote locations. 


The recent COVID-19 pandemic saw a rise in demand for portable and rapid diagnostic systems \cite{Ahmad2020}. There have been first experiments towards compact imaging systems that have attracted attention for applications beyond the laboratory \cite{Schaefer2012}. Metasurfaces, defined as thin photonic devices with sub-wavelength structure over a 2D surface, have emerged as promising solutions \cite{Neshev2023}. Flat meta-lenses are being integrated into portable devices \cite{Dirdal2022} and other metasurfaces are finding use in areas including phase imaging \cite{Wesemann2021, Wang2023} and multi-modal microscopy \cite{Intaravanne2023, Badloe2023, Wu2023}. For example, a metasurface was used to synchronously generate spiral phase contrast and bright field images \cite{Zhang2023}, albeit requiring two redirected beams with additional optical components. Recent developments have also centered on adding tunability to these devices \cite{Ling2023}, leveraging techniques such as phase-change materials \cite{Rogers2016, Yang2021, Khodasevych2023, Tripathi2021, Cotrufo2024}, liquid crystals \cite{Zografopoulos2015}, opto-fluidics \cite{vandeGroep2022}, mechanical straining \cite{Zhou2021} and electrical gating \cite{Abdollahramezani2022}. As an external parameter within an imaging system, polarization also presents an attractive approach to optical tunability \cite{Sharma2023,Cheng2018}. It has proved instrumental for dynamically controlling color filtering \cite{Vashistha2017}, lensing \cite{Ou2022} and phase imaging \cite{Engay2021,Huo2020}.


Here, we present the first demonstration of a metasurface designed for polarization-tunable, near-infrared phase contrast and bright field multi-modal microscopy. The metasurface consists of rectangular blocks of silicon on a glass substrate. When an incident electric field has a component parallel to the long axis of the blocks, the metasurface generates phase contrast images via a Mie resonance. However, bright field images are produced under the orthogonal polarization, enabling bi-modal switching with a switching contrast of up to 93\%. Importantly, this does not require interchanging components into and out of the system nor the use of beam-splitters, as it is usually the case in multi-modal microscopy methods. We experimentally demonstrate this with \textit{Caenorhabditis elegans} samples, as well as unstained human prostate cancer and breast tissue samples. Pathologists commonly use chemical stains, such as haematoxylin and eosin (H\&E) \cite{OToole2021}, to image these samples because imaging breast tissue, for example, can be difficult due to small ($\Delta n \sim 10^{-2})$ refractive index difference relative to the surrounding \cite{Parvin2021}. However, staining is an irreversible process that introduces additional costs and can delay the turnover time in pathology. We show that our proposed imaging method can be used to distinguish between breast ducts and blood vessels in unstained breast tissue derived from a patient with breast cancer. Moreover, the use of near-infrared illumination offers deeper penetration into tissue compared to visible light \cite{Qiu2018}, while retaining the use of low-cost silicon photo-detectors. The dual approach of polarization switchability and infrared illumination offers an advance in miniaturized multi-modal microscopy. It holds the potential to deliver prompt results where existing technologies are unavailable and aligns with the need for compact imaging systems in point-of-care settings \cite{Pham2013}.


\section*{Results}\label{section:results}


\subsection*{Tunable multi-modal microscopy - principles}\label{subsection:mutlimodal-img-princ}

The fundamental concept for the application of the metasurface to tunable multi-modal microscopy is the filtering of the spatial frequencies of an image through object-plane image processing. The optical transfer function of the metasurface, defined as the Fourier transform of its Green's function, characterizes its response to the spatial frequencies of an input image. In general, it is a rank-2 tensor given in the $\{p,s\}$-polarization basis as,
\begin{equation}
    \label{eq:otf-tensor}
    H(k_x,k_y)= \begin{pmatrix}
    H_{pp}(k_x,k_y) & H_{ps}(k_x,k_y) \\
    H_{sp}(k_x,k_y) & H_{ss}(k_x,k_y) \\ \end{pmatrix} \ .
\end{equation}
Henceforth, we utilize an $\{x,y\}$ basis, i.e. $H_{xx}$, $H_{xy}$, $H_{yx}$ and $H_{yy}$, describing the impact of the metasurface on the transverse components of the incident field. By taking the $z$-axis as the optical axis, $k_x$ and $k_y$ denote the transverse spatial frequency components of the wave-vector $\vec{k}$. Ignoring vector effects for now for illustration, the impact of the metasurface on the spatial frequencies of an input image of a semi-transparent sample $O(x,y) = O_0(x,y) e^{i \varphi(x,y)}$, where $O_0$ is the amplitude and $\varphi(x,y)$ is the phase, can be modelled using the convolution theorem, 
\begin{equation}
    \label{eq:otf-conv-thm}
    I(x,y) \approx \left| \mathscr{F}^{-1} \left\{ H(k_x, k_y) \Tilde{O}(k_x, k_y) \right\} \right|^2 \ .
\end{equation}
Here, $I$ is the output intensity image, $\mathscr{F}$ denotes the Fourier transform and $\Tilde{O} = \mathscr{F} \{O\}$. Note that $|O(x,y)|^2$ is the bright field (absorption) image that is independent of $\varphi(x,y)$. If we let the $x$-axis be aligned with the long axis of the silicon blocks for an un-tilted metasurface (normal incidence), then the metasurface has an approximately linear optical transfer function at a tilt of $\SI{2}{\degree}$ about the $y$- (short) axis (Fig. \ref{fig:metasurface}c) for $x$-polarized illumination, i.e. $H_{xx} \propto k_x + k_{x0}$, where $k_{x0}$ is a constant. This can compute the derivative along the $x$-direction, plus a constant offset that retains a non-zero background in the output image. In the case of a pure phase sample $|O(x,y)| \propto O_0 = \text{constant}$, the sample is visualized because the phase derivatives are converted into intensity variations, e.g.
\begin{equation}
    \label{eq:phase-gradient}
    I(x,y) \approx \left|O_0 \left( \frac{\partial \varphi(x,y)}{\partial x} + k_{x0} \right) \right|^2 \ .
\end{equation}
One could also uniquely distinguish the phase derivatives as they will manifest as different intensity levels that are above or below the background (region of zero phase gradient), i.e. $I(x,y) \approx |O_0 k_{x0}|^2$. On the other hand, the metasurface is approximately transparent when the illumination is $y$-polarized, which corresponds to a flat optical transfer function, $H_{yy} \approx \text{constant}$. In this case, the output image is proportional to the bright field image, $I(x,y) \propto |O(x,y)|^2$. Therefore, with this configuration, $x$-polarized illumination leads to the production of phase contrast images, while $y$-polarized illumination leads to bright field images. It is shown in the Supplementary Information S2.1 that the effects of cross-polarization are negligible, i.e. $H_{xy}$ and $H_{yx}$ in this case are zero, and we ignore diffraction effects.

We define the switching contrast of the metasurface as $\delta = H_{yy}(0,0) - H_{xx}(0,0)$ and the numerical aperture (NA) of the metasurface as the width of the optical transfer function in $k_x$ where the above conditions are met. This is in contrast to the numerical aperture of the imaging system in which the metasurface is used. Other performance metrics of the metasurface are given in the Supplementary Information S2.2, as well as further details of the theoretical principles (S1.1) and polarization used in this work (S3.1).


\subsection*{Design and optical response}


\begin{figure*}[!t]
   \centering
    \includegraphics[width=\linewidth]{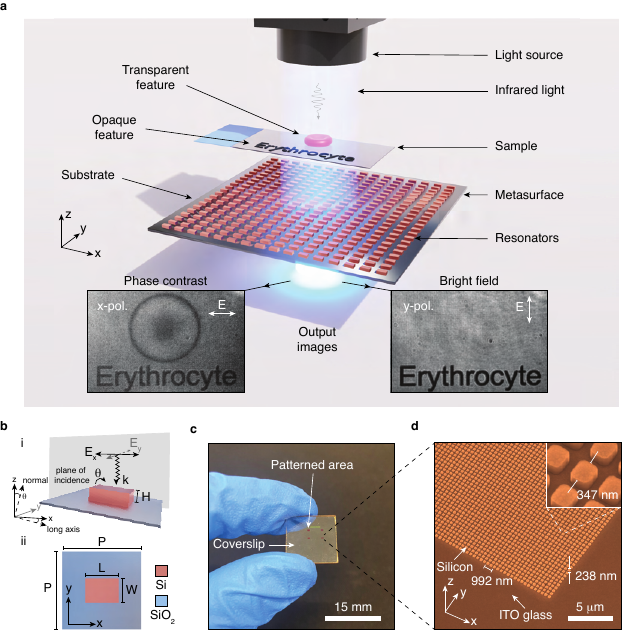}
    \caption{\textbf{Design of the metasurface.} (a) A stylized depiction of use of the metasurface to perform tunable multi-modal microscopy, with experimental images of a mock red blood cell (erythrocyte) given in the insets. (b) The composition of the metasurface shown in 3D (metasurface tilted) (i) and bird's eye (metasurface un-tilted) (ii) views of the unit cell. A schematic of the polarization is also given in i. (c) The fabricated coverslip with patterned regions of the metasurface. (d) Scanning electron microscope images of the metasurface.}
    \label{fig:metasurface}
\end{figure*}

A schematic of the approach used here is given in Fig. \ref{fig:metasurface}a. The metasurface consists of a 2D lattice of rectangular silicon resonators of length $L = \SI{322}{\nano\meter}$, width $W = \SI{293}{\nano\meter}$ and height $H = \SI{236}{\nano\meter}$ (Fig. \ref{fig:metasurface}b). The unit cell of the structure is repeated with a lattice constant of $P = \SI{526}{\nano\meter}$ in both directions (Fig. \ref{fig:metasurface}b-ii). The fabricated metasurface was patterned on a glass wafer \SI{15}{\milli\meter} by \SI{15}{\milli\meter} in size (Fig. \ref{fig:metasurface}c) and the total size of each metasurface was \SI{760}{\micro\meter} by \SI{507}{\micro\meter}. The fabrication steps are outlined in the \hyperref[section:methods]{Methods} section. Scanning electron microscope images of the fabricated metasurface reveal the rectangular geometry of the resonators (Fig. \ref{fig:metasurface}d and Supplementary Information S4.1), which is the basis of the polarization tunability.


\begin{figure*}[!ht]
   \centering
    \includegraphics[width=\linewidth]{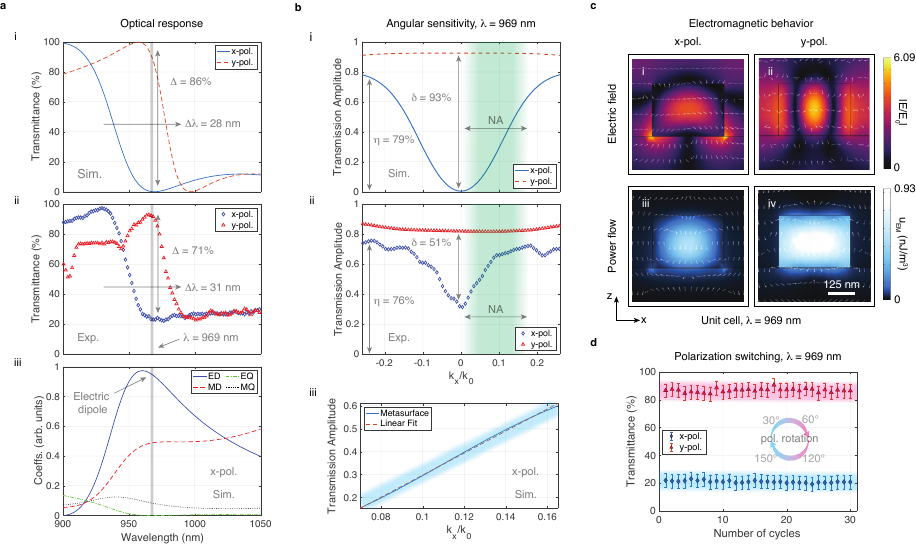}
    \caption{\textbf{Optical response of the metasurface.} ((a)-i) The simulated normal-incidence transmission spectrum of the metasurface for $x$- and $y$-polarization. ((a)-ii) Same as a-i but for experiment. ((a)-iii) The simulated multipole decomposition of the normal-incidence transmission spectrum under $x$-polarization. The electric dipole resonance is indicated by the grey shaded region. ((b)-i) The magnitude of the simulated co-polarized components of the optical transfer function along $k_y = 0$ at the operating wavelength for $x$- and $y$-polarization. ((b)-ii) Same as b-i but for experiment. ((b)-iii) Linear fitting of the optical transfer function in b-i. (c) The simulated magnitude (heat map) and direction (arrows) of the electric field (top), and the electromagnetic energy density (heat map) and time-averaged Poynting vector (arrows) (bottom) in the $xz$-plane of the unit cell. (d) The experimental normal-incidence transmittance through the metasurface as the polarization was switched between $x$ and $y$.}
    \label{fig:optical_response}
\end{figure*}

The optical response of the metasurface was investigated by measuring its angle-dependent transmission. Simulations were carried out using a commercial electromagnetic solver (COMSOL Multiphysics 6.1) implementing the finite element method, which is detailed in the \hyperref[section:methods]{Methods} section. The experiments were performed using the normal-incidence and angle-dependent spectroscopy configurations described in the \hyperref[section:methods]{Methods} section and in the Supplementary Information S3.2. The simulated (Fig. \ref{fig:optical_response}a-i) and experimental (Fig. \ref{fig:optical_response}a-ii) normal-incidence transmission spectra of the metasurface show a minimum in the transmission at a wavelength of \SI{969}{\nano\meter} under $x$-polarized illumination. This set the operating wavelength to be $\lambda = \SI{969}{\nano\meter}$. The transmission minimum reached a level of 0.002\% in the simulation and 20\% in the experiment. It also red-shifted by $\Delta \lambda \approx \SI{30}{\nano\meter}$ in both the simulation and experiment when the polarization was switched to $y$-polarization. This produces a polarization-switchable transmission at the operating wavelength, with a contrast of $\Delta = T_y - T_x = 86\%$ in the simulation and $\Delta = 71\%$ in the experiment. Here, $T_x$ and $T_y$ are the transmittance levels under $x$- and $y$-polarized illumination, respectively. By using the method of Ref. \cite{Terekhov2019} detailed in the Supplementary Information S1.2, a multipole decomposition of the simulated transmission spectrum under $x$-polarization (Fig. \ref{fig:optical_response}a-iii) revealed that the transmission minimum was associated with an electric dipole Mie resonance, which is discussed further in the Supplementary Information S1.3. 

Plots of the electric field (Fig. \ref{fig:optical_response}c-top) and power flow (Fig. \ref{fig:optical_response}c-bottom) at the operating wavelength show the different responses of the metasurface. Under $x$-polarization, the electric dipole excitation preferentially reflects the illumination, which is apparent in the power flow and in the normal-incidence reflectance spectrum given in the Supplementary Information S2.1. On the other hand, the power flow is directed through the metasurface under $y$-polarization. The quality factor of the resonance was calculated to be $Q = 17.3$ in the simulation and $Q = 19.3$ in the experiments, which are of the order of the typical values reported for Mie resonances \cite{Koshelev2021}. Two tables are given in the Supplementary Information (S2.2 and S4.3) that summarize the various performance metrics of the metasurface. Furthermore, the electromagnetic behavior of the metasurface at the operating wavelength in all of the cross-sectional planes of the unit cell are also given in the Supplementary Information S2.1.

The angle-dependent transmission through the metasurface as a function of the polar angle of incidence $\theta$ was simulated and experimentally measured. Here, $\theta$ represents the angle from the optical axis ($z$-axis) in the $xz$-plane (Fig. \ref{fig:metasurface}c-i), corresponding to an azimuthal angle of $\phi = \SI{0}{\degree}$. Using the relationship between the angles of incidence of plane waves and the spatial frequencies of light that is explained in the Supplementary Information S1.1, the angle-dependent transmission amplitude through the metasurface is equivalent to the magnitude of the relevant component of the optical transfer function \cite{Davis2021}. The results (Fig. \ref{fig:optical_response}b) show the magnitude of the co-polarized components of the optical transfer function along $k_y=0$ as a function of the normalized spatial frequency $k_x/k_0$, where $k_0 = 2\pi/\lambda$ is the wavenumber. 

For $x$-polarization, the component $|H_{xx}(k_x,0)|$ exhibits a high-pass lineshape with a processing efficiency of $\eta = \max(|H_{xx}(k_x^{\text{max}},0)|) = 79\%$ in the simulation (Fig. \ref{fig:optical_response}b-i) and $\eta = 76\%$ in the experiment (Fig. \ref{fig:optical_response}b-ii). Here, $k_x^{\text{max}}$ is the maximum spatial frequency component at the edge of the `contrast zone', which we define as the shaded region in Fig. \ref{fig:optical_response}b-i-ii where the metasurface can be used for tunable multi-modal microscopy. It has a maximum numerical aperture of $\text{NA} \approx 0.17$. Linear fitting of the simulated optical transfer function within the range $k_x/k_0 = 0.07$ to $0.17$ is shown in Fig. \ref{fig:optical_response}b-iii. The linear fit, derived using the polyfit function and curve fitting toolbox in MATLAB, can be represented by the relation $(4.8 \pm 0.11) k_x/k_0 - (0.18 \pm 0.014)$. On the other hand, the component $|H_{yy}(k_x,0)|$ was approximately flat with $k_x$ under $y$-polarization in both the simulation (Fig. \ref{fig:optical_response}b-i) and experiment (Fig. \ref{fig:optical_response}b-ii). It exhibited a switching contrast of $\delta = 93\%$ in the simulation and $\delta = 51\%$ in the experiment. This difference was due to the experimental transmittance level being 20\% higher than the simulated level at normal incidence under $x$-polarization, which corresponds to a 40\% transmission amplitude difference. Finally, the experimental switching of the metasurface at the operating wavelength (Fig. \ref{fig:optical_response}d) showed consistent performance as the linear polarization of the illumination was repeatedly switched between $x$ and $y$. Additional simulation data is given in the Supplementary Information S2, including the full optical transfer function tensor and transmission spectra as functions of the polarization, the refractive index of the surrounding and the geometric parameters of the metasurface.
 

\subsection*{Tunable multi-modal microscopy - experiment}

\begin{figure*}[!ht]
   \centering
    \includegraphics[width=\linewidth]{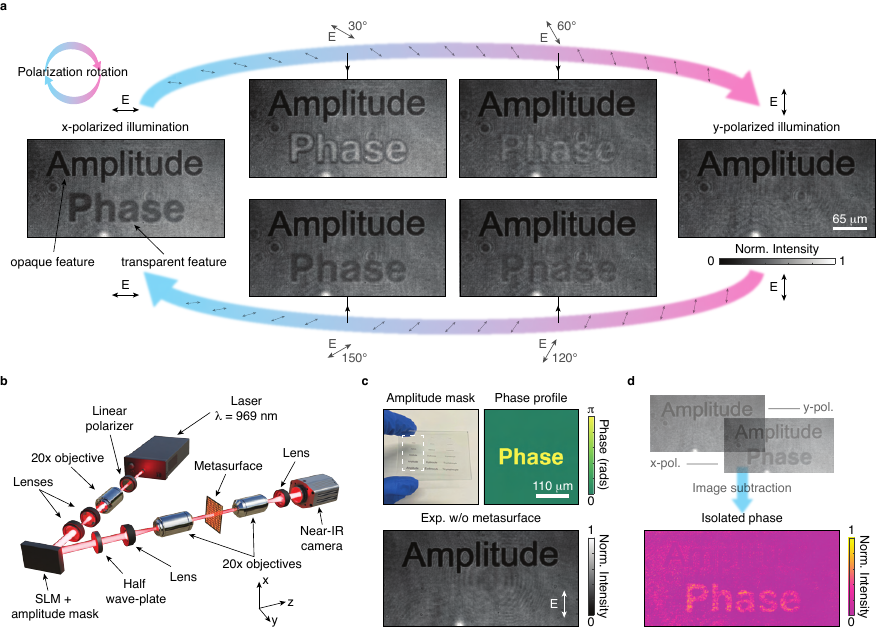}
    \caption{\textbf{Tunable multi-modal microscopy of samples generated by a spatial light modulator.} (a) A flowchart of the images produced by the metasurface for different polarization states of the illumination. (b) The configuration used in the experiment. (c-top left) The amplitude mask. (c-top right) The phase profile shown on the SLM. (c-bottom) The image taken in the absence of the metasurface. (d) The image obtained by subtracting the image acquired under $x$-polarization from that acquired under $y$-polarization.}
    \label{fig:slm_imaging}
\end{figure*}


Tunable multi-modal microscopy experiments were conducted on a range of samples using the metasurface and near-infrared illumination at the operating wavelength of \SI{969}{\nano\meter}. In all of the measurements, the metasurface was tilted by \SI{2}{\degree} to the optical ($z$) axis, corresponding to a normalized spatial frequency of $k_{x0}/k_0 \approx 0.03$, about the short axis of the resonators (illustrated in Fig. \ref{fig:metasurface}c and the Supplementary Information S3.1). This accessed the contrast zone of the metasurface by introducing an offset to the optical transfer function to become the shifted $k$-space origin about the angular offset. Consequently, some of the background associated with low spatial frequencies was retained in the output images, similar to that seen in DIC. This produced the appearance of pseudo-3D contrast on a non-zero background. All of the output images were normalized with respect to their brightest pixels. Finally, the imaging experiments provided a field of view of approximately $\SI{500}{\micro\meter}$ and a resolution limit of $\rho \geq 0.5 \lambda/\text{NA} = \SI{2.85}{\micro\meter}$. 


The first set of experiments, shown in Fig. \ref{fig:slm_imaging}b and detailed in the \hyperref[section:methods]{Methods} section, involved a computer-controlled spatial light modulator (SLM) that replicated the phase profiles of a range of transparent samples with a phase excursion of $\pi$ radians. This included the word `Phase' (Fig. \ref{fig:slm_imaging}c-top right) that was next to a purely absorptive amplitude mask containing the word `Amplitude' (Fig. \ref{fig:slm_imaging}c-top left) placed directly in front of the SLM. The steps taken to fabricate the amplitude mask are detailed in the \hyperref[section:methods]{Methods} section. The other phase profiles that were used on the SLM included models of a human red blood cell (Fig. \ref{fig:metasurface}a) and a leukaemic Jurkat cell, which were each next to the words `Erythrocyte' and `T-Lymphocyte' from the amplitude mask, respectively. The imaging results of the cells are presented in the Supplementary Information S2.4, as well as simulations of the imaging experiments.

The images produced by the metasurface were organized in a flowchart with respect to the different polarization states of the illumination (Fig. \ref{fig:slm_imaging}a). Both of the transparent (`Phase') and opaque (`Amplitude') features were visible in the image obtained under $x$-polarization (Fig. \ref{fig:slm_imaging}a-left). The contrast seen of the word `Phase' transitioned from a maximum to a minimum as the polarization was rotated from $x$ to $y$ (Fig. \ref{fig:slm_imaging}a-top). The opposite effect was observed when the polarization was rotated from $y$ to $x$ (Fig. \ref{fig:slm_imaging}a-bottom). It was also observed that the contrast of the word `Phase' flipped between the two cases. This was a consequence of accessing either the positive or negative slope of the optical transfer function. Meanwhile, the word `Amplitude' retained a similar appearance in all of the images and was the only feature visible in the images obtained under $y$-polarization (Fig. \ref{fig:slm_imaging}a-right) and in the absence of the metasurface (Fig. \ref{fig:slm_imaging}c-bottom). Finally, the image obtained under $x$-polarization was subtracted from that obtained under $y$-polarization (Fig. \ref{fig:slm_imaging}d) to qualitatively isolate the phase. However, parts of the word `Amplitude' remained visible in the subtracted image that can be attributed to some non-uniformity between the two images.


\begin{figure*}[!ht]
   \centering
    \includegraphics[width=\linewidth]{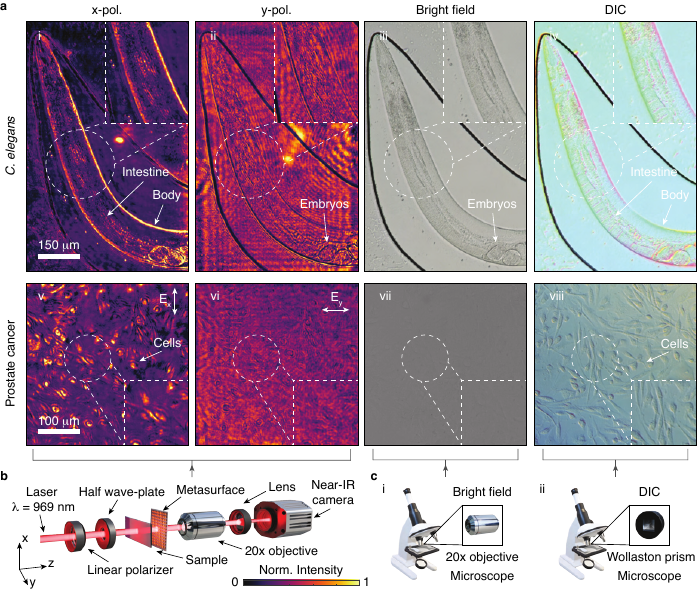}
    \caption{\textbf{Tunable multi-modal microscopy of biological samples.} (a) The output images of \textit{C. elegans} (i-iv) and human prostate cancer cells (v-viii). These include images obtained with $x$-polarized illumination (first column, false-color), $y$-polarized illumination (second column, false-color), a bright field microscope (third column, true color) and a DIC microscope (final column, true color). (b) A schematic of the experimental configuration. (c) Schematics of bright field (i) and DIC (ii) microscopes. The dashed lines in the images indicate the regions that are zoomed in the insets.}
    \label{fig:bio_imaging}
\end{figure*}

To demonstrate the potential of the metasurface for bio-imaging applications, tunable multi-modal microscopy was experimentally performed on samples of \textit{C. elegans} (Fig. \ref{fig:bio_imaging}a-i-iv) and unstained human prostate cancer cells (Fig. \ref{fig:bio_imaging}a-v-viii). The experiments were performed using the configuration illustrated in Fig. \ref{fig:bio_imaging}b. The steps taken to prepare the samples and details of the experimental configuration are given in the \hyperref[section:methods]{Methods} section. For comparison, bright field (Fig. \ref{fig:bio_imaging}a-third column) and DIC (Fig. \ref{fig:bio_imaging}a-last column) microscope images were obtained in the absence of the metasurface using the configurations illustrated in Fig. \ref{fig:bio_imaging}c.

The semi-transparent body and embryos of the \textit{C. elegans} were visible in the image obtained under $x$-polarization (Fig. \ref{fig:bio_imaging}a-i), similar to the DIC microscope image (Fig. \ref{fig:bio_imaging}a-iv). On the other hand, only the embryos and the outline of the body were visible in the images obtained under $y$-polarization (Fig. \ref{fig:bio_imaging}a-ii) and with the bright field microscope (Fig. \ref{fig:bio_imaging}a-iii). Two videos of the imaging experiments are provided in the Supplementary Information, including a recording of a live nematode, demonstrating the capacity for real-time dynamic monitoring. The images of the prostate cancer cells had similar distinguishing features between the different types of imaging modalities. The cellular structure and internal details were visible in the image obtained under $x$-polarization (Fig. \ref{fig:bio_imaging}a-v) and in the DIC microscope image of a similar region of the sample (Fig. \ref{fig:bio_imaging}a-viii). On the other hand, only some of the outlines of the cells were visible under $y$-polarization (Fig. \ref{fig:bio_imaging}a-vi) and in the bright field microscope image (Fig. \ref{fig:bio_imaging}a-vi).


\begin{figure*}[!ht]
   \centering
    \includegraphics[width=\linewidth]{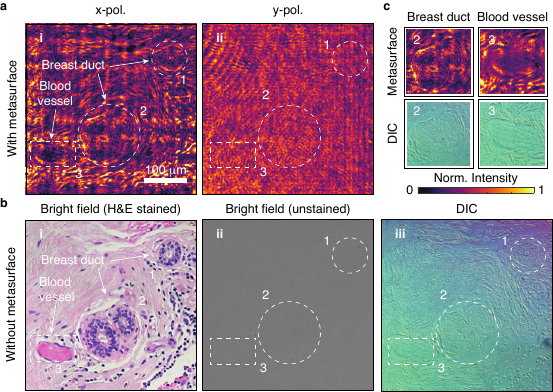}
    \caption{\textbf{Tunable multi-modal histology of human breast tissue.} (a) The images obtained with the metasurface under $x$- ((a)-i, false-color) and $y$-polarization ((a)-ii, false-color). ((b)-i) A bright field microscope image of the sample stained with H\&E obtained under unpolarized white light illumination (true color). ((b)-ii) The same as (b)-i but for the unstained sample (true color). ((b)-iii) A DIC microscope image of the unstained sample obtained under white light illumination (true color). (c) The insets of the numbered regions in the images in (a)-i and (b)-iii.}
    \label{fig:histology}
\end{figure*}

Finally, the metasurface was applied to imaging unstained human breast tissue using the configuration illustrated in Fig. \ref{fig:bio_imaging}b, as before. The samples were derived from retrospective patients diagnosed with ductal carcinoma in situ during the course of regular pathology. The steps taken to prepare the samples are detailed in the \hyperref[section:methods]{Methods} section. For comparison, an additional bright field microscope image was taken of the sample stained with H\&E (Fig. \ref{fig:histology}b-i). 

As anticipated, nothing was visible in the images of the unstained sample with the metasurface under $y$-polarization (Fig. \ref{fig:histology}a-ii) and with a bright field microscope (Fig. \ref{fig:histology}b-ii). The bright field microscope image of the stained sample (Fig. \ref{fig:histology}b-ii) displayed clear and color-coded contrast. The image contained normal breast tissue elements, including background fibrotic stroma, small capillaries (Fig. \ref{fig:histology}b-i, feature 3), mature lymphocytes and normal breast ducts (Fig. \ref{fig:histology}b-i, features 1 and 2) with a dual cell layer involving myo-epithelial cells and inner epithelial cells with central secretion material. Some of these elements can be identified in the images produced by the metasurface under $x$-polarization (Fig. \ref{fig:histology}a-i) and a DIC microscope (Fig. \ref{fig:histology}b-iii). In particular, the simple circular outline of the small capillary can be identified at feature 3 and three benign breast ducts can be identified as simple toroidal structures at features 1 and 2. These normal architectural patterns in tissue can then be distinguished from tumors that often form aberrant arrangements. However, the capacity to visualize finer architectural complexity or cellular detail is uncertain, as the region only included relatively simple benign structures and individual cellular elements could not be visualized using these methods. The sample was significantly thicker than the samples in Fig. \ref{fig:bio_imaging}, which provided a challenge for the imaging modalities. A more ideal application would be the preparation of cytology material from either a touch or smear preparation, which could provide thinner samples and allow for assessing individual cellular elements in the images.


\section*{Discussion}\label{section:discussion}


Many multi-modal microscopy systems typically involve complicated configurations due to the use of beam-splitters, fiber bundles or the multiplexing of wavelengths. While the cascading of imaging systems can generally be implemented in a laboratory, it can be difficult to integrate into portable systems. The metasurface in this article could generate bi-modal images within the same beam and field of view and has the potential to be integrated into a portable system that is considerably smaller than existing configurations. It did not require the use of beam-splitters or dichroic mirrors and was limited only by the speed of polarization switching. The control over the imaging modality through the polarization is a relatively straightforward technique given that polarization is an external parameter of an imaging system. It can be adjusted either mechanically or electronically, such as through the use of electronic variable retarders. The use of a polarization-sensitive camera could further enable single-shot capturing of the bi-modal images. 


The image quality produced by the metasurface was similar to that achieved in DIC microscopy (for $x$-polarization) and bright field microscopy (for $y$-polarization). In DIC microscopy, the image quality heavily relies on the optical path length differences not exceeding a specific threshold to avoid destructive interference that can lead to reduced visibility or artifacts in regions of strong phase gradients. On the other hand, the metasurface relies only on the spatial frequencies of the input image overlapping with the numerical aperture of the metasurface. It is also unlike other Fourier-based techniques such as Zernike microscopy or Schlieren imaging, which often struggle to discriminate between positive and negative phase gradients of an optical field and tend to produce artifacts in images when the phase gradients are too strong. What sets the metasurface apart is that it could produce images without the need for contrast-enhancing agents, nor the use of digital processing or scanning commonly associated with computational techniques such as the transport of intensity equation \cite{Li2016} or ptychography \cite{Gao2017}. However, phase gradients corresponding to spatial frequency information that is outside the numerical aperture of the metasurface can be missed, therefore requiring magnification or a larger numerical aperture.


Dielectric metasurfaces, particularly those with high refractive index materials such as the silicon in this study, are useful due to their durability and strong Mie resonances \cite{Komar2021}. The metasurface presents opportunities for integration into existing portable systems, including smart phones, in combination with LEDs and artificial intelligence to classify images. The use of meta-lenses can further reduce the overall size of the system \cite{Khorasaninejad2016}. The simple geometry of the metasurface holds potential for scalable mass production and can serve as building blocks for more complicated structures optimized through an inverse design approach. While the relatively small numerical aperture of the metasurface is advantageous for imaging larger samples, it can potentially miss finer details or smaller samples without further magnification. A metasurface with an enhanced or tunable numerical aperture could overcome this. The anisotropy of the 2D optical transfer function of the metasurface (Supplementary Information S2.1) was a consequence of the use of rectangular resonators. Therefore, precise rotations of the metasurface with respect to the optical axis were required to ensure consistent performance.


The findings in this article highlight the potential of the metasurface to be used in a wide range of applications, including real-time biological imaging and dynamic monitoring \cite{Kang2011}. Specifically, it has potential for the early detection of structural abnormalities in tissue without the requirement for time-consuming and costly sample processing, nor expensive optical microscopes. It emerges as a potentially valuable tool for health professionals that can be used in conjunction with other techniques for point-of-care testing where access to skilled laboratory labour and specialized hardware may be unavailable. Finally, some possible extensions to this work could encompass 2D image processing \cite{Wang2022}, combining other imaging modalities such as fluorescence microscopy or polarization-difference imaging, and hyperspectral imaging due to the sensitivity of the metasurface to the refractive index of the surrounding.  

\section*{Conclusions}\label{section:conclusion}

In conclusion, we have demonstrated a polarization-tunable, near-infrared multi-modal microscopy of biological and other samples using a metasurface. The metasurface exhibited an approximately linear optical transfer function for polarization along one direction due to the excitation of a Mie resonance, while maintaining an approximately flat optical transfer function for the orthogonal polarization. This property enabled the bi-modal switching between phase contrast and bright field imaging with a switching contrast of up to $\delta = 93\%$. It was experimentally demonstrated across a range of samples, including \textit{C. elegans}, prostate cancer cells and breast tissue. Notably, we distinguished between breast ducts and blood vessels in unstained breast tissue derived from a patient with breast cancer. Overall, the importance of these results lies in their capacity to address a gap where other techniques are either unavailable or require complex sample preparation. It offers an advance toward the development of compact imaging technology, holding promising applications in point-of-care testing and remote pathology \cite{Nguyen2022}. 


\subsection*{Methods}\label{section:methods}


\subsubsection*{Fabrication}

The fabrication of the metasurface started with the deposition of \SI{236}{\nano\meter} of amorphous silicon using plasma-enhanced chemical vapor deposition (PECVD, Oxford Instruments PLASMALAB 100 PECVD) on a \SI{23}{\nano\meter} film of indium tin oxide (ITO) coating a \SI{10}{\milli\meter} thick glass substrate. This was then spin-coated with approximately \SI{200}{\nano\meter} of poly-methyl methacrylate resist (MicroChem, PMMA 950A) and electron beam lithography (Vistec EBPG5000plusES) patterned the structure with a dosage of \SI{900}{\micro\coulomb\per\centi\meter\squared}. This was followed by a step of electron beam evaporation (Intlvac Nanochrome II) of \SI{10}{\nano\meter} of aluminium oxide (Al$_2$O$_3$) and a step of lift-off to create the hard mask for silicon etching. Pseudo Bosch etching was performed by inductively coupled plasma - refractive ion etching (ICP-RIE, Oxford Instruments PlasmaPro System100). The scanning electron microscope images were taken using a scanning electron microscope (FEI NovaNanoSEM 430). A flowchart of the fabrication steps is given in the Supplementary Information S4.1. 


\subsubsection*{Simulations}

Numerical simulations were performed using COMSOL Multiphysics 6.1 implementing the finite element method within the wave optics module. Periodic Floquet boundary conditions were applied to the sides of the unit cell of the metasurface to model an infinite 2D lattice. The glass substrate and air superstrate were assumed to be homogeneous and lossless materials with a refractive index of 1.5 and 1, respectively. They were both taken to be \SI{750}{\micro\meter} thick, which was much thicker than the silicon resonators. The silicon material was assumed to be a homogeneous and lossless material with a refractive index of 3.67 at the wavelength of $\lambda = \SI{969}{\nano\meter}$. Port boundary conditions were applied onto the top and bottom of the unit cell to launch and absorb electromagnetic plane waves, respectively, and to monitor the amplitude and phase of reflected and transmitted orders. The plane waves had an electric field strength of \SI{1}{\volt/\meter} and an incident power of \SI{1}{\watt}. The finite meshing elements in the model had a maximum size of $\lambda/$\SI{8}{\nano\meter} and a minimum size of \SI{5}{\nano\meter}. The optical transfer function of the metasurface was obtained by computing the complex coefficient $S_{21}$ of the scattering matrix of the two-port system.


\subsubsection*{Normal-incidence spectroscopy}

The metasurface was mounted on an XYZ stage and was illuminated by collimated white light from a fiber-coupled halogen lamp (Mikropack HL-2000-FHSA) through a microscope objective (Olympus UPlanFI 4x/0.13). The illumination was linearly polarized (Thorlabs LPVIS050-MP2 550nm-1.5um) to produce $x$- or $y$-polarized light. The light transmitted through the metasurface was collected by a microscope objective (Nikon ELWD 50x/0.55) and a lens (Thorlabs lens LA1131-A $f = \SI{50}{\milli\meter}$). A second lens (Thorlabs lens LA1951-A $f = \SI{25}{\milli\meter}$) focused the output light onto the fiber-port of a spectrometer (Ocean Insight NIR quest). The acquisition time of each measurement was \SI{1}{\sec} and the data was normalized with respect to the spectra obtained from an unpatterned region of the sample (i.e. the glass substrate). A beam-splitter (Thorlabs CM1-BS013) was used before the second lens to relay an image of the metasurface onto a camera (Thorlabs DCC1645C-HQ) using a lens (Thorlabs lens LA1608-B f = 75 mm). A schematic of the experimental configuration is given in the Supplementary Information S3.2. 


\subsubsection*{Angle-dependent spectroscopy}

The angle-dependent spectroscopy experiments were performed using a back focal plane imaging configuration. The metasurface was illuminated with a broadband tungsten halogen white light source (Ocean Optics HL-2000-HP) through an air microscope objective (Olympus 10x/0.25). The metasurface was mounted on the stage of an inverted microscope (Olympus IX71). The illumination was linearly polarized (Thorlabs LPVIS100) to produce $x$- or $y$-polarized light. It was then expanded to fill the microscope objective back aperture to excite the metasurface with the full angular range as determined by the numerical aperture of the objective. The reflected light over the same angular range was collected by the same microscope objective and an image of the back focal plane was relayed to a spectrograph (Acton SP300i)/CCD camera (Princeton Pixis 1024BR) couple by a tube lens (Olympus single port tube lens $f = \SI{180}{\milli\meter}$) and Bertrand lens (Thorlabs TTL180-A $f = \SI{180}{\milli\meter}$) pair. A second polarizer (Thorlabs LPVIS100) selected the co- or cross-polarized components of the output. One axis of the back focal plane image was transformed into direction by the spectrograph to provide an angle-resolved reflectance distribution about the wavelength of \SI{969}{\nano\meter}. The acquisition time of each measurement was \SI{90}{\sec} and the data was normalized with respect to the data obtained using a mirror in place of the metasurface. The normalized reflectance data was converted to transmittance data by $T = 1 - R - A$, where the absorptance $A$ of the metasurface was calculated to be approximately zero in the Supplementary Information S2.1-2.2. The transmission amplitude was then obtained by taking the square root of the transmittance data. A schematic of the experimental configuration is given in the Supplementary Information S3.2.


\subsubsection*{Multi-modal microscopy - spatial light modulator}

The imaging experiments shown in Fig. \ref{fig:slm_imaging} were performed using light from a super-continuum source (NKT Photonics SuperK COMPACT laser). The wavelength was controllable through the NKT Photonics Control software and a tunable multi-channel filter (NKT SuperK SELECT). The power of the light was measured to be \SI{0.02}{\milli\watt} by a power meter (Thorlabs PM1000D). The illumination was linearly polarized (Thorlabs LPNIR050-MP2) to be compatible with the SLM (Holoeye Pluto-VIS-001 LCOS-SLM). The reflected light was expanded and collimated by a microscope objective (Olympus Plan N 20x/0.40) and two lenses (Thorlabs LA1027-B $f = \SI{35}{\milli\meter}$ and LA1509-B $f = \SI{100}{\milli\meter}$) to fill the window range of the SLM. The amplitude imaging mask was mounted directly in front of the SLM and the SLM replicated phase profiles from a computer with a phase excursion of $\pi$ radians. A half wave-plate (Thorlabs WPH05M-1064) was used to produce either $x$- or $y$-polarized light. A lens (Thorlabs LA1433-B $f = \SI{150}{\milli\meter}$) and microscope objective (Nikon LU Plan 20x/0.4) projected a de-magnified image of the SLM onto the metasurface. The metasurface was tilted by \SI{2}{\degree} using an XYZ rotation stage. A microscope objective (Olympus Plan N 20x/0.4) and a lens (Thorlabs LA1986-B $f = \SI{125}{\milli\meter}$) relayed the magnified output image onto a camera (Andor Zyla 4.2P sCMOS). The exposure time of the camera was \SI{0.06}{\sec}. A schematic of the experimental configuration is given in the Supplementary Information S3.3.

The amplitude imaging mask was \SI{50}{\milli\meter} by \SI{40}{\milli\meter} in size. It was fabricated on a \SI{2.5}{\milli\meter} thick soda lime glass photo-mask plate with \SI{150}{\nano\meter} of chrome (Cr) and an anti-reflective coating. The plate was coated with AZ1518 photo-resist and it was exposed to ultra-violet light with a wavelength of \SI{365}{\nano\meter} (IMP600 direct write system, dose \SI{75}{\milli\joule\per\centi\meter\squared}) to define the pattern. The exposed photo-resist was removed by developing in AZ400K developer diluted in water (4 parts water to 1 part AZ400K) for \SI{1}{\minute}. It was then rinsed with de-ionized water and dried with nitrogen gas (N$_2$). The exposed chrome was removed by immersing the plate in a chrome etchant (TechniEtch CR01) for \SI{60}{\sec} and then it was rinsed with de-ionized water and dried with nitrogen gas (N$_2$). The remaining resist was removed by exposing it to ultra-violet light with a wavelength of \SI{365}{\nano\meter} (IMP600 direct write system, dose \SI{150}{\milli\joule\per\centi\meter\squared}) and developing it in AZ400K developer diluted with water. 


\subsubsection*{Multi-modal microscopy - biological samples and polarization switching}

The biological imaging experiments (Figs. \ref{fig:bio_imaging} and \ref{fig:histology}) were performed using light from a super-continuum source (NKT Photonics SuperK COMPACT and NKT Photonics SuperK SELECT), as above. It was linearly polarized (Thorlabs LPNIR050-MP2) to produce $x$- or $y$-polarized light. This illuminated the biological sample that was mounted on an XYZ sample holder (Thorlabs SFH3 fixed filter holder). A half wave-plate (Thorlabs WPH05M-1064) was used to control the polarization. The metasurface was placed immediately after the sample and it was tilted by \SI{2}{\degree} by an XYZ rotation stage. A microscope objective (Nikon LU Plan 20x/0.4) and a lens (Thorlabs LA1986-B $f = \SI{125}{\milli\meter}$) relayed the output image onto a camera (Andor Zyla 4.2P sCMOS). A schematic of the experimental configuration is given in the Supplementary Information S3.4.

The polarization switching experiment (Fig. \ref{fig:slm_imaging}d) was performed with the same configuration. In this case, the metasurface was illuminated with $x$- or $y$-polarized light at normal incidence and there was no imaging sample placed in the experiment. The output transmittance through the metasurface was measured using the camera. The transmittance was normalized to that measured in the absence of the metasurface. The measurements were repeated for different states of linear polarization that was controlled by the half wave-plate.


\subsubsection*{Sample preparation}

\textbf{\textit{C. elegans}}. The \textit{C. elegans} were wild type (stain N2) animals and were obtained from the Caenorhabditis Genetics Center at the University of Minnesota. The cultures were maintained at \SI{20}{\celsius} using standard protocols on nematode growth media (NGM) supplement with \textit{E. coli} (strain OP50) \cite{Brenner1974}. Developmentally synchronous animals that were derived from eggs laid over a \SI{2}{\hour} window, and then cultured for an additional \SI{72}{\hour}, were used for imaging. These young adult worms were isolated, mounted under glass cover-slips on 2\% (w/v) agarose pads, and immobilized via brief heat shock of \SI{60}{\celsius} for \SI{10}{\sec} prior to imaging. 

\textbf{Prostate cancer cells}. The human prostate cancer cells (PC-3) were obtained from the American Type Cell Collection (ATCC, USA). The cells were seeded on a glass coverslip at a seeding density of 300,000 cells. The cells were then grown in complete RPMI media (10\% FBS, 1\% Penicillin/streptomycin) for \SI{24}{\hour} in a humidified incubator at \SI{37}{\celsius} with 5\% carbon dioxide (CO$_2$). The cells were then washed with Dulbecco's phosphate-buffered saline (DPBS) and fixed with 4\% paraformaldehyde. The glass coverslip with the cells was mounted on a microscope slide. The RPMI 1640 Culture medium, fetal bovine serum, Penicillin/streptomycin antibiotic, Dulbecco's phosphate-buffered saline (no calcium, no magnesium) and 16\% methanol free paraformaldehyde, were all procured from ThermoFisher Scientific, Australia.

\textbf{Breast tissue}. The human breast tissue samples were obtained surgically through the course of regular pathology and were prepared at Dorevitch Pathology in Melbourne. The samples were obtained with informed consent from the patients that was approved by the University of Melbourne ethics approval (project ID 27303). The tissue was fixed with formalin and contained known invasive breast tumors, for which all clinical diagnostic sampling and formal reporting had been completed. The tissue samples were approximately 3-\SI{4}{\milli\meter} thick and were taken from benign and tumor-involved regions of the breast. These sections underwent standard laboratory tissue processing with paraffin infusion on the Evident VIP6 platform. The samples were then embedded within paraffin into tissue blocks and cut via microtomy into sections that were \SI{3}{\micro\meter} in thickness. These tissue slices were then placed onto charged microscope slides. One of the tissue sections was stained with haematoxylin and eosin staining profile and a coverslip was applied.


\subsubsection*{Conventional microscopy}

Bright field and DIC microscopy were performed using an inverted microscope (Olympus BX60) with white light illumination, a microscope objective (Olympus U Plan FI 20x/0.5) and a camera (Nikon Z5). A green filter (Grubb Parsons \SI{590}{\nano\meter} filter) was used in DIC microscopy to enhance the contrast of the output image.


\backmatter

\bmhead{Supplementary information}

Supplementary material is available in the Supplementary Information. This includes one Supplementary Information document and four Supplementary videos. 

\bmhead{Authors' contributions}

A.R, S.B.S and L.W conceptualized the project. The idea of near-infrared phase microscopy with a silicon-on-glass metasurface was put forward by A.R. S.B.S designed the metasurface and put forward the idea of polarization-tunable multi-modal microscopy with L.W. S.B.S conducted the theoretical and computational work. The metasurface was fabricated by J.M at the Melbourne Centre for Nanofabrication, after preliminary versions of the metasurface were fabricated by L.W. N.P built the optical setup for normal-incidence spectroscopy and J.A.H built the optical configuration for angle-dependent spectroscopy. S.B.S collected the normal-incidence spectroscopy data and also the angle-dependent spectroscopy data with assistance from J.A.H. S.B.S built the optical configuration for imaging and performed the imaging experiments. W.S prepared the prostate cancer samples and G.M prepared the \textit{C. elegans} samples. K.R obtained the breast tissue specimens, which were prepared as fixed pathology samples by M.M and Dorevitch Pathology in Melbourne. M.M provided a pathology analysis of the breast tissue images and contributed to the discussion of the images. S.B.S processed the data, created the figures, wrote the article and prepared the manuscript. All authors provided input on the manuscript. A.R, L.W, K.B.C and A.A.S supervised the project.

\bmhead{Acknowledgments}

This work was performed in part at the Melbourne Centre for Nanofabrication (MCN) in the Victorian node of the Australian National Fabrication Facility (ANFF). Dan Smith from MCN is acknowledged for preparing the imaging target used in the experiments involving the spatial light modulator. Wendy S. L. Lee from the University of New South Wales and Dragomir Neshev from the Australian National University are acknowledged for their support in preliminary spectroscopy measurements. Rajour T. Ako from RMIT University is acknowledged for contributing to the etching process in the fabrication of a preliminary version of the metasurface. Ravi Shukla, also from RMIT University, is acknowledged for providing PC-3 cell culture resources. The authors further acknowledge Timothy J. Davis, Timur Gureyev, Snjezana Tomljenovic-Hanic, Sivacarendran Balendhran and Benjamin Russell from the University of Melbourne for their useful discussions.  

\subsection*{Declarations}

\bmhead{Funding}

This research was funded by the Australian Government through the Australian Research Council (ARC) Centre of Excellence grant CE200100010. S.B.S acknowledges the support of the Ernst \& Grace Matthaei Scholarship and the Australian Government Research Training Program Scholarship. J.A.H acknowledges the ARC projects CE170100026 and FT180100295. N.P acknowledges the support of a Melbourne Research Scholarship. 

\bmhead{Competing interests}

The authors declare no competing interests. 

\bmhead{Ethics approval}

Ethics approval and informed patient consent for the breast tissue samples were approved by the University of Melbourne Human Research Ethics Committee under the application identification number 27303.

\bmhead{Availability of data and materials}

The data that support the findings within this paper are available from the corresponding authors upon reasonable request.

\bmhead{Code availability}

The computer code that support the findings within this paper is available from the corresponding authors upon reasonable request. 

\bibliography{References.bib}


\setcounter{section}{0}
\setcounter{figure}{0}
\renewcommand{\thesection}{S\arabic{section}}
\renewcommand\thefigure{S\arabic{figure}}
\renewcommand{\theequation}{S\arabic{equation}}

\newpage

\vspace{3em}
\begin{center}
    {\Titlefont Metasurfaces for infrared multi-modal microscopy: phase contrast and bright field - supplementary information \par}
\end{center}

\vspace{3em}
\tableofcontents{}
\newpage 



\section{Theory}

\subsection{Image processing}

The fundamental concepts for the application of the metasurface to tunable multi-modal microscopy are briefly described in the main article. Here, the explanation is expanded to explain how the metasurface operates in the formation of images. Consider a source of monochromatic plane waves with an electric field $\Vec{E}_s(x,y,z)$, wavelength $\lambda$ and wave-vector $\Vec{k}$ propagating along the $z$-axis. Assuming polarized light and ignoring vector effects for the time being, suppose that this source of light illuminates a semi-transparent sample located in the $z=0$ plane with a transmission function $O(x, y) = O_0(x,y) e^{i \varphi(x, y)}$ in the Kirchhoff approximation. Here, $O_0(x,y)$ represents the amplitude features of the sample and $\varphi(x,y)$ represents the phase features. This defines an optical image in the $z=0$ plane with an intensity profile given by $I_0(x,y) = |O(x,y)|^2 = |O_0(x,y)|^2$, which represents the bright field image of the sample. Visualizing the phase features of the sample requires highlighting the phase shifts imparted onto the light transmitted through the sample, which can be modelled as,
\begin{equation}
\label{eq:sample-field}
    \Vec{E}_{ \text{in} }(x,y) = O_0(x,y) e^{i \varphi(x, y)} \Vec{E}_s(x,y) \ .
\end{equation}
The metasurface filters the spatial frequency components of an image in either the image plane of the sample or directly in the object plane. The optical response of the metasurface in $k$-space is characterized by the optical transfer function, which is defined as the Fourier transform of its Green's function. In general, the optical transfer function is a rank-2 tensor (to accommodate polarization) given in the $\{p,s\}$-polarization basis as,
\begin{equation}
    \label{eq:otf-tensor}
    H(k_x,k_y)= \begin{pmatrix}
    H_{pp}(k_x,k_y) & H_{ps}(k_x,k_y) \\
    H_{sp}(k_x,k_y) & H_{ss}(k_x,k_y) \\ \end{pmatrix} \ .
\end{equation}
The spatial frequencies $k_x$, $k_y$ and $k_z = \sqrt{ |\Vec{k}|^2 - k_x^2 - k_y^2}$ are the Cartesian projections of $\Vec{k}$, where $k_0 = |\Vec{k}| = 2 \pi/\lambda$ is the wavenumber. As in Ref. \cite{Davis2021SI}, the spatial frequency components can be represented as a function of the spherical coordinates $(r = k_0, \theta, \phi)$,
\begin{equation}
    \label{eq:spatial_freqs_angles}
    \begin{gathered}
        k_x = k_0 \sin(\theta) \cos(\phi) \\
        k_y = k_0 \sin(\theta) \sin(\phi) \\
        k_z = k_0 \cos(\theta) \ ,
    \end{gathered}
\end{equation}
where $(\theta, \phi)$ are the polar and azimuthal propagation angles of the plane waves with respect to the $z$-axis, respectively. Any light beam can be decomposed into bundles of weighted plane waves, each travelling in different directions, by the Fourier transform \cite{Goodman1996SI},
\begin{equation}
    \label{eq:fourier_transform}
    \mathscr{F} \{E(x,y)\}(k_x,k_y) = \int^\infty_{-\infty} \int^\infty_{-\infty} E(x,y) e^{-i (k_x x + k_y y)} dx dy \ ,
\end{equation}
where $\mathscr{F}$ is the Fourier transform operator. Also, given that the optical transfer function represents the response of the metasurface to plane waves in $k$-space, then it follows that the spatial frequency dependence of the metasurface in $k$-space is equivalent to an angular dependence in real space. This is the concept that is fundamental to the design of the metasurface. 

The impact of the metasurface on an input image is determined by the Green's function of the metasurface. The output image is given by a convolution integral between the input image and the Green's function,
\begin{equation}
    \label{eq:convolution}
    E_i(x,y) = \int_{-\infty}^\infty \int_{-\infty}^\infty G_{ij}(x - x', y - y') E_j^{\text{in}}(x',y')  dx' dy' \ ,
\end{equation}
where $G(x,y)$ is the Green's function tensor and $E_i$ is the $i^{th}$-component of the output electric field. However, by the convolution theorem, the impact of the metasurface on the spatial frequency spectrum of the input image can be directly modelled as,
\begin{equation}
    \label{eq:otf-conv-thm-SI}
    \Tilde{E}_i(k_x,k_y) = H_{ij}(k_x, k_y) \Tilde{E}^{\text{in}}_j(k_x, k_y) \ ,
\end{equation}
where $\Tilde{E} = \mathscr{F} \{E\}$. The convolution theorem simplifies the relationship between the input image and the output image from a convolution in real space to a multiplication with the optical transfer function in $k$-space. The output electric field $E$ can be recovered from $\Tilde{E}$ by the inverse Fourier transform, 
\begin{equation}
    \label{eq:inverse_fourier_transform}
    \mathscr{F}^{-1} \{\Tilde{E}(k_x, k_y)\}(x,y) = \frac{1}{2\pi} \int^\infty_{-\infty} \int^\infty_{-\infty} \Tilde{E}(k_x, k_y) e^{i (k_x x + k_y y)} dk_x dk_y \ ,
\end{equation}
where $\mathscr{F}^{-1}$ is the inverse Fourier transform operator and $E = \mathscr{F}^{-1}\{\Tilde{E}\}$. Therefore, given the following property for the Fourier transform of a derivative, 
\begin{equation}
    \label{eq:FT-derivatives}
    \mathscr{F} \left\{ \frac{\partial E}{\partial x} \right\} (k_x, k_y) = i k_x \Tilde{E}(k_x, k_y) \ ,
\end{equation}
then a linear optical transfer function, $H \propto k_x$, can produce the spatial derivative (up to a multiplicative constant) of an incoming light field along the $x$-direction. It follows that a metasurface with an approximately linear angular dispersive transmission $T(\theta) \propto \sin(\theta) \approx \theta$ (for $\phi = 0$ and small values of $\theta$) is capable of phase visualization in the object or image plane of a sample. Here, the \{$p$, $s$\}-polarization basis was used but, in general, the optical transfer function tensor can be expressed in any polarization basis. Henceforth, we utilize an $\{x,y\}$ basis, i.e. $H_{xx}$, $H_{xy}$, $H_{yx}$ and $H_{yy}$, describing the impact of the metasurface on the transverse components of the incident field. It was shown in the main article that if we let the $x$-axis be aligned with the long axis of the silicon blocks for an un-tilted metasurface (normal incidence), then the metasurface has an approximately linear optical transfer function at a tilt of $\SI{2}{\degree}$ about the $y$- (short) axis for $x$-polarized illumination, i.e. $H_{xx} \propto k_x + k_{x0}$, where $k_{x0}$ is a constant. This can compute the derivative along the $x$-direction, plus a constant offset that pertains to a non-zero background in the output image. Hence, the design of the metasurface was carried out by optimizing the angular response of the metasurface in real space to engineer its optical transfer function in $k$-space. 

In the case of a pure phase sample that has $|O_0(x,y)| = \text{constant}$, then the sample is visualized because the phase derivative is converted into intensity variations in the output image $I(x,y)$,
\begin{equation}
    \label{eq:phase-gradient-SI}
    I(x,y) \approx \left|O_0 \left( \frac{\partial \varphi(x,y)}{\partial x} + k_{x0} \right)\right|^2 \ .
\end{equation}
This result is obtained when a linear optical transfer function ($H_{xx} \propto k_x + k_{x0}$) and Eq. \eqref{eq:sample-field} are substituted into Eq. \eqref{eq:otf-conv-thm-SI}. Hence, the output image is proportional to \(| \partial \varphi(x,y)/\partial x |^2\), with contrast created in regions where the phase is varying along the direction of differentiation. The constant offset in the optical transfer function is due to the metasurface being slightly tilted by \SI{2}{\degree}, corresponding to a normalized spatial frequency of $k_{x0}/k_0 \approx 0.03$, that shifts the $k$-space origin about the angular offset. The increased transmission level at this offset preserves some of the illumination to create a non-zero background in the output image. This distinguishes the imaging modality from dark field imaging that has a dark background in the output image. One could also uniquely distinguish the phase derivatives as they will manifest as different grayscale levels that are above or below the background, corresponding to the region of zero phase gradient, i.e. $I(x,y) \approx |O_0 k_{x0}|^2$. 

On the other hand, the metasurface was approximately transparent when the illumination was $y$-polarized. This corresponds to a flat optical transfer function, $H_{yy} \approx \text{constant}$, that only attenuates the input image. In the case that the constant is equal to one, the output image will be exactly the input image. As a result, the output image is proportional to the bright field image, i.e. $I(x,y) \propto |O(x,y)|^2$. The cross-polarized components of the optical transfer function $H_{xy}$ and $H_{yx}$ (Fig. \ref{fig:2d_otf} and Table \ref{tbl:performance_sim}) are zero in the region of operation. However, they have some non-zero contributions for larger values of $k_x$ and $k_y$, but these contributions were measured to be negligible in the experiments by using an analyzing polarizer on the output image. 

The performance of the polarization switching was quantified by the switching contrast of the metasurface, defined as $\delta = H_{yy}(0,0) - H_{xx}(0,0)$. A similar value can be defined for the switching contrast of the transmittance, given by $\Delta = T_y - T_x$, where $T_x$ and $T_y$ are the transmittance levels under $x$- and $y$-polarized illumination at the operating wavelength, respectively. The efficiency of the metasurface for image processing is defined as $\eta = |H_{xx}(k_x^{\text{max}},0)|$, where $k_x^{\text{max}}$ is the maximum spatial frequency component where the optical transfer function of the metasurface is still within its numerical aperture. The numerical aperture of the metasurface, NA, is defined to be the width of the optical transfer function where the above conditions are met. This is contrasted from the numerical aperture of the imaging system itself, which is determined by the optical components in the system. 


\subsection{Multipole decomposition}

Fig. 2a-iii of the main article shows a multipole decomposition of the transmittance spectrum of the metasurface. The method from Ref. \cite{Terekhov2019SI} was used to do this and it is briefly summarized here. Consider a metasurface composed of an array of identical dielectric resonators. For normally incident, $x$-polarized illumination, the multipole decomposition of the transmittance of the metasurface up to second order is given by, 
\begin{equation}
    \label{eq:multipole}
        T = \left | 1 + \frac{ik}{2 E_0 A_{\text{cell}} \epsilon_0 \epsilon_s} \left ( p_x + \frac{1}{v_s} m_y - \frac{ik}{6} Q_{xz} - \frac{ik}{2v_s} M_{yz} + \cdots\right ) \right |^2 \ .
\end{equation}
Here, $k = 2\pi/\lambda$ is the wave-number in the surrounding, $E_0$ is the magnitude of the incident electric field, $\lambda$ is the wavelength of the illumination, $A_{\text{cell}}$ is the area of the lattice unit cell, $\epsilon_s$ is the relative permittivity of the surrounding and $v_s$ is the speed of light in the surrounding. Furthermore, $\Vec{p}$ is the electric dipole moment vector, $\Vec{m}$ is the magnetic dipole moment vector, $Q$ is the electric quadrupole moment tensor and $M$ is the magnetic quadrupole moment tensor. The relevant components of the multipole moments are given by,
\begin{equation}
    \label{eq:multipole_moments}
    \begin{gathered}
        p_x = \int P_x j_0(kr) dV + \frac{k^2}{10} \int \frac{15 j_2(kr)}{(kr)^2} \left( (\Vec{r} \cdot \Vec{P})x - \frac{1}{3} r^2 P_x \right) dV \\
        m_y = - \frac{i \omega}{2} \int \frac{3 j_1(kr)}{kr} \left( \Vec{r} \times \Vec{P} \right)_y dV \\
        Q_{xz} = \int \frac{9 j_1(kr)}{kr} \left( x P_z + P_x z \right) dV + 6k^2 \int \frac{j_3(kr)}{(kr)^3} \left( 5xz (\Vec{r} \cdot \Vec{P}) - (x P_z + P_x z) r^2 \right) dV \\
        M_{yz} = \frac{\omega}{3i} \int \frac{15 j_2(kr)}{(kr)^2} \left( (\Vec{r} \times \Vec{P})_y z + y (\Vec{r} \times \Vec{P})_z \right) dV \ ,
    \end{gathered}
\end{equation}
where $ \Vec{r} = (x,y,z)$ is the position vector, $j_n$ is the $n^{th}$-order spherical Bessel function of the first kind, $\Vec{P}(\Vec{r}) = \epsilon_0 (\epsilon_s - 1) \Vec{E}(\Vec{r})$ is the polarization density and $\omega = 2 \pi f$ is the angular frequency. The integrals are performed over the volume $V$ of the resonator within a single unit cell of the metasurface. This formalism was set up in COMSOL Multiphysics 6.1 to obtain the multipole decomposition of the transmission spectrum of the metasurface.


\subsection{Mie scattering of an isolated resonator}

In 1908, Gustav Mie \cite{Mie1908SI} exactly solved Maxwell's equations for the scattering of plane waves from a homogeneous sphere of size comparable to a wavelength. Although the analytic solution exists only for ellipsoidal particles, the physics is identical for arbitrary geometries \cite{Bohren1983SI}. Consider the Mie scattering of plane waves by a rectangular resonator. To characterize its multipolar behaviour, the electric field scattered by the resonator when under plane wave illumination is expanded in the basis of vector spherical harmonics, 
\begin{equation}
    \label{eq:multipole-expansion}
    \Vec{E}_{sc}(r, \theta, \phi) = \sum_{n=1}^\infty \sum_{m=-n}^{+n} c_{nm} \left( a_{nm} \Vec{N}_{nm}(r,\theta,\phi) + b_{nm} \Vec{M}_{nm}(r, \theta, \phi) \right) \ ,
\end{equation}
where $a_{nm}$ and $b_{nm}$ are the electric and magnetic scattering coefficients, respectively. The normalization constants are defined as,
\begin{equation}
    \label{eq:norm-const}
    c_{nm} = |\Vec{E}_0| i^{n+2m-1} \sqrt{ \frac{(2n + 1) (n-m)!}{4 (n+m)!}} \ ,
\end{equation}
where $\Vec{E}_0$ is the electric field of the illumination. Following the formalism from Ref. \cite{Muhlig2011SI}, the electric and magnetic harmonics are respectively defined in spherical coordinates as,
\begin{equation}
    \label{eq:VSH}
    \begin{gathered}
    \Vec{N}_{nm} = \frac{1}{k_0 r} \biggl( \tau_{nm}(\cos(\theta)) \Hat{\theta} + i \pi_{nm}(\cos(\theta)) \Hat{\phi} \biggr) e^{im\phi} \frac{\partial}{\partial r} \biggl( r h_n^{(1)}(k_0 r) \biggr) \\
    + \frac{n(n+1)}{k_0 r} P_n^m(\cos(\theta)) h_{n}^{(1)}(k_0 r) e^{im\phi} \Hat{r} \\
    \Vec{M}_{nm} = h_n^{(1)}(k_0 r) e^{im\phi} \biggl( i \pi_{nm}(\cos(\theta)) \Hat{\theta} - \tau_{nm}(\cos(\theta)) \Hat{\phi} \biggr) \ ,
    \end{gathered}
\end{equation}
where $h_n^{(1)}$ is the $n^{th}$-order spherical Hankel function of the first-kind and $P_n^m$ is the $n^{th}$-order associated Legendre polynomial of degree $m$. The $\tau$ and $\pi$ functions are defined as,
\begin{equation}
    \label{eq:tau-pi}
    \begin{gathered}
        \tau_{nm}(\theta) = \frac{\partial}{\partial \theta} P_n^m(\cos(\theta)) \\
        \pi_{nm}(\theta) = m P_n^m(\cos(\theta)) \csc(\theta) \ .
    \end{gathered}
\end{equation}
The scattering coefficients can be computed by projecting the scattered electric field onto the vector spherical harmonics on a virtual sphere of radius $r_0$ surrounding the resonator, given by,
\begin{equation}
    \label{eq:scat-coeffs}
    \begin{gathered}
        a_{nm} = \frac{ \int_0^{2\pi} \int_0^\pi \Vec{E}_{sc}(r=r_0, \theta, \phi) \cdot \Vec{N}^*_{nm}(r=r_0, \theta, \phi) \sin(\theta) d\theta d\phi }{ c_{nm} \int_0^{2\pi} \int_0^\pi | \Vec{N}_{nm}(r=r_0, \theta, \phi) |^2 \sin(\theta) d\theta d\phi } \\
        b_{nm} = \frac{ \int_0^{2\pi} \int_0^\pi \Vec{E}_{sc}(r=r_0, \theta, \phi) \cdot \Vec{M}^*_{nm}(r=r_0, \theta, \phi) \sin(\theta) d\theta d\phi }{ c_{nm} \int_0^{2\pi} \int_0^\pi | \Vec{M}_{nm}(r=r_0, \theta, \phi) |^2 \sin(\theta) d\theta d\phi } \ .
    \end{gathered}
\end{equation}
By using the asymptotic relation $h_n^{(1)}(k_0 r) \sim (-i)^n e^{ik_0 r}/ik_0 r$ for $k_0 r \gg 1$ and the orthonormality of the associated Legendre polynomials, it follows that the scattering cross section is given by
\begin{equation}
    \label{eq:cross-sect}
    \sigma_{s} = \frac{\pi}{k_0^2} \sum_{n=1}^\infty \sum_{m=-n}^{+n} n(n+1) \left( |a_{nm}|^2 + |b_{nm}|^2 \right) \ .
\end{equation}
The integer $n \in [1,\infty)$ provides the degree of the multipole, i.e. $n = 1$ corresponds to dipole, $n = 2$ to quadrupole, etc., while the integer $m \in [-n, n]$ provides their orientations in space. The contribution to the scattering cross section for each $n$ represents the cross section for that particular degree of the multipole. Furthermore, for each multipole degree, the electric component corresponds to the $a_{nm}$ term and the magnetic component to the $b_{nm}$ term. Hence, the total cross section can be decomposed as \cite{Fruhnert2017SI},
\begin{equation}
    \label{eq:cross-sect-multipole}
    \sigma_{s} = \sigma_s^{ED} + \sigma_s^{MD} + \sigma_s^{EQ} + \dotsm \ ,
\end{equation}
where ED, MD and EQ each stand for electric dipole, magnetic dipole and electric quadrupole, respectively. The Mie resonances of the resonator manifest as peaks in the cross sections at the resonant wavelengths. This formalism was set up in COMSOL Multiphysics 6.1 to determine the multipolar decomposition of the scattering cross section for an isolated resonator from the metasurface.  


\section{Additional simulation data}

\subsection{Optical response}

\begin{figure*}[!t]
   \centering
    \includegraphics[width=0.9\linewidth]{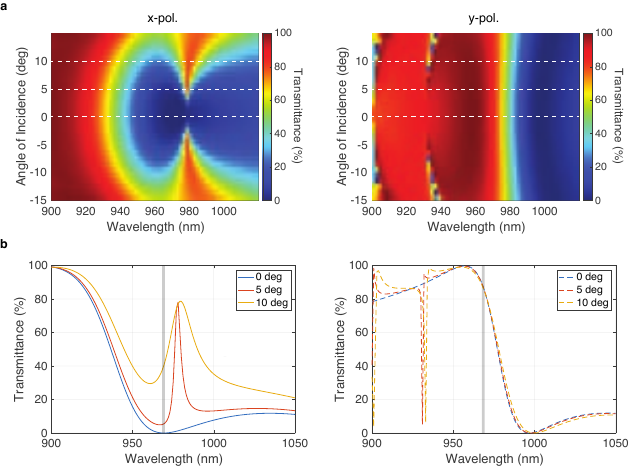}
    \caption{\textbf{Angle-dependent transmission spectra}. (a - left) A 2D heat map of the transmission versus the angle of incidence and the wavelength of $x$-polarized illumination. (a - right) Same as the left but for $y$-polarization. (b - left) The 1D transmission spectra for 0, 5 and 10 degrees angle of incidence for $x$-polarized illumination. (b - right) Same as the left but for $y$-polarized illumination.}
    \label{fig:angle_T}
\end{figure*}

Simulations of the optical response of the metasurface were performed using the steps outlined in the Methods section of the main article. The metasurface was designed to have an angle-dependent transmission for $x$-polarized illumination at the operating wavelength, and an angle-independent transmission for $y$-polarized illumination. The angle-dependent transmission spectra are given in Fig. \ref{fig:angle_T} as a function of the angle of incidence. The results show that the transmission spectrum changed under $x$-polarized illumination with a varying angle of incidence. Meanwhile, the transmission spectrum under $y$-polarized illumination remained similar when the angle of incidence was varied, with only the appearance of dark modes for shorter wavelengths. 

\begin{figure*}[!t]
   \centering
    \includegraphics[width=0.9\linewidth]{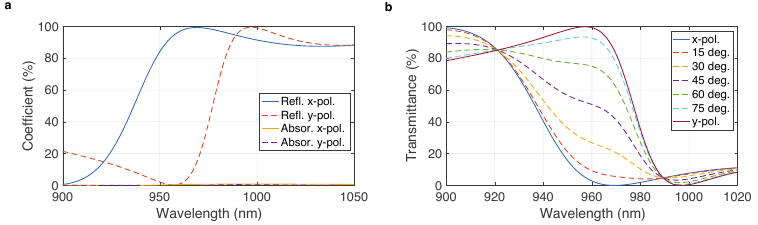}
    \caption{\textbf{1D spectra as a function of various parameters}. (a) The reflectance and absorptance spectra of the metasurface for $x$- and $y$-polarized illumination. (b) The transmittance spectra as a function of the angle of linear polarization.}
    \label{fig:1D_spectra}
\end{figure*}

The 1D plots of the optical response are given in Fig. \ref{fig:1D_spectra}. The reflectance of the metasurface is shown to be a maximum at the operating wavelength for $x$-polarized illumination and a minimum for $y$-polarized illumination (Fig. \ref{fig:1D_spectra}a). The absorptance is also shown to be nearly zero for both cases. The transmission spectra for different states of linear polarization shows how the spectrum morphs between those for $x$- and $y$-polarized illumination (Fig. \ref{fig:1D_spectra}b).

\begin{figure*}[!t]
   \centering
    \includegraphics[width=0.9\linewidth]{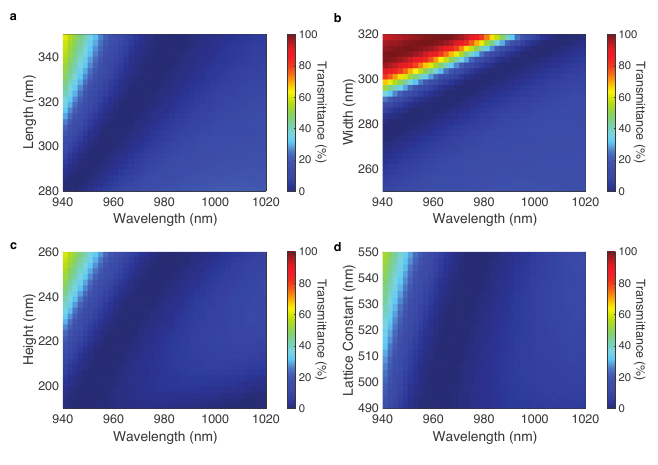}
    \caption{\textbf{Dispersion of the resonance}. (a) A 2D heat map of the transmittance spectrum as a function of the length of the resonators. (b) Same as a but for the width of the resonators. (c) Same as a but for the height of the resonators. (d) Same as a but for the lattice constant of the metasurface. In all cases, the illumination was $x$-polarized.}
    \label{fig:dispersion}
\end{figure*}

The operating wavelength of the metasurface is dependent on the geometrical parameters of the metasurface, including the size of the resonators (length L, width W and height H) and the lattice constant (P) of the array. The dispersion of the resonance as a function of the geometrical parameters is shown in Fig. \ref{fig:dispersion}. In each case, the resonance red-shifted as the geometrical parameters were increased. 

\begin{figure*}[!t]
   \centering
    \includegraphics[width=0.9\linewidth]{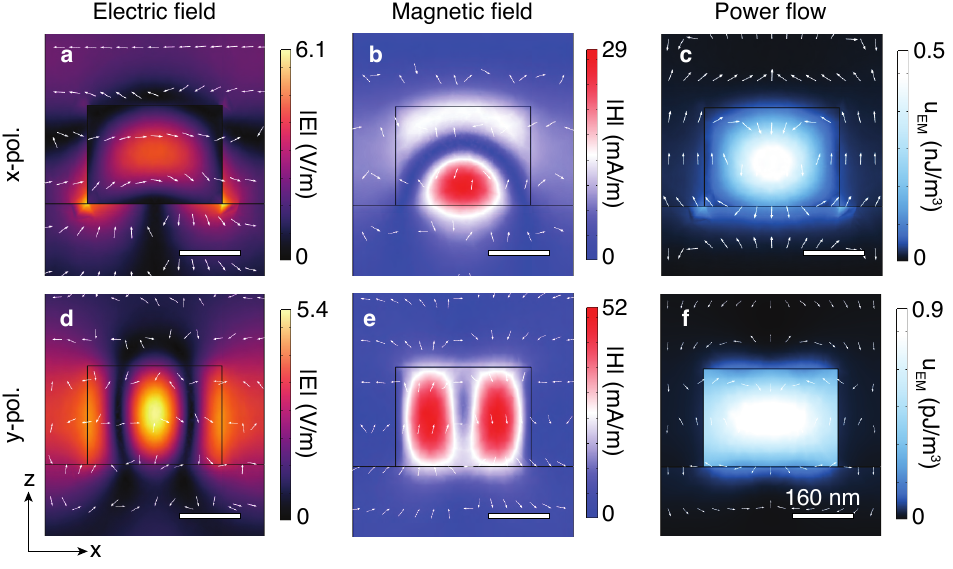}
    \caption{\textbf{Electromagnetic behaviour of the metasurface in the $xz$-plane}. (a),(d) The electric field strength (heat map) and direction (arrows) in the unit cell for $x$- (a) and $y$-polarization (d). (b),(e) Same as (a),(d) but for the magnetic field. (c),(f) The time-averaged electromagnetic energy density (heat map) and Poynting vector (arrows) in the unit cell for $x$- (c) and $y$-polarization (f). All of the images are given in the $xz$-plane of the metasurface and at the operating wavelength.}
    \label{fig:fields_xz}
\end{figure*}

\begin{figure*}[!t]
   \centering
    \includegraphics[width=0.9\linewidth]{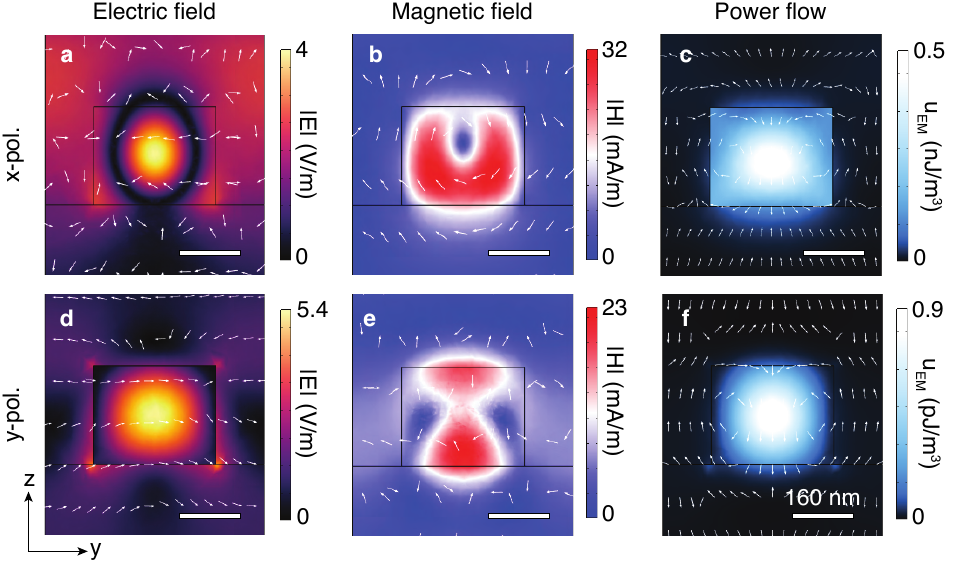}
    \caption{\textbf{Electromagnetic behaviour of the metasurface in the $yz$-plane}. (a),(d) The electric field strength (heat map) and direction (arrows) in the unit cell for $x$- (a) and $y$-polarization (d). (b),(e) Same as (a),(d) but for the magnetic field. (c),(f) The time-averaged electromagnetic energy density (heat map) and Poynting vector (arrows) in the unit cell for $x$- (c) and $y$-polarization (f). All of the images are given in the $yz$-plane of the metasurface and at the operating wavelength.}
    \label{fig:fields_yz}
\end{figure*}

\begin{figure*}[!t]
   \centering
    \includegraphics[width=0.9\linewidth]{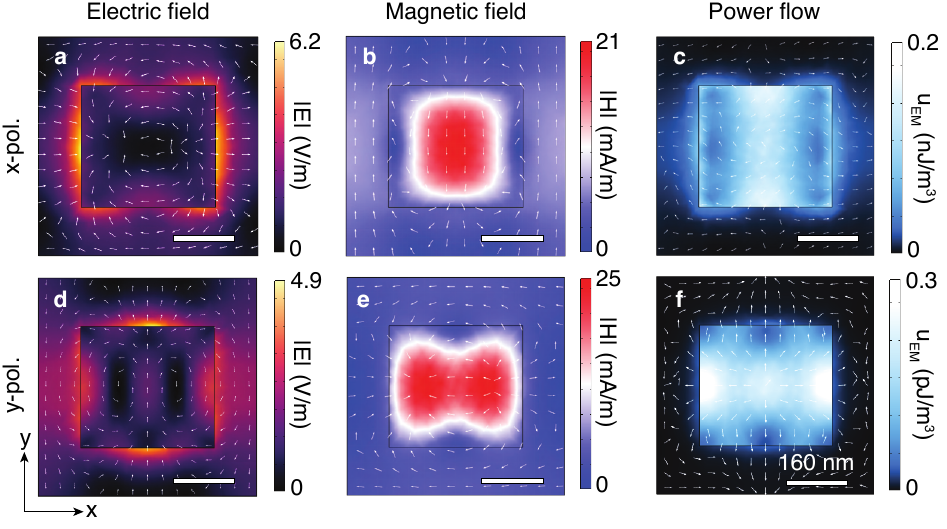}
    \caption{\textbf{Electromagnetic behaviour of the metasurface in the $xy$-plane}. (a),(d) The electric field strength (heat map) and direction (arrows) in the unit cell for $x$- (a) and $y$-polarization (d). (b),(e) Same as (a),(d) but for the magnetic field. (c),(f) The time-averaged electromagnetic energy density (heat map) and Poynting vector (arrows) in the unit cell for $x$- (c) and $y$-polarization (f). All of the images are given in the $xy$-plane of the metasurface and at the operating wavelength.}
    \label{fig:fields_xy}
\end{figure*}

The electromagnetic behavior of the metasurface when on-resonance is depicted in Figs. \ref{fig:fields_xz}-\ref{fig:fields_xy}. They each show the electric field, magnetic field and power flow in the unit cell of the metasurface at the operating wavelength for $x$- and $y$-polarized illumination. The electric and magnetic fields when on-resonance from $x$-polarized illumination display behavior that is linked to an electric dipole resonance, with the power of the fields swirling around the resonator. On the other hand, the fields behave differently when off-resonance and the power of the fields flows through the resonator. 

\begin{figure*}[!t]
   \centering
    \includegraphics[width=0.9\linewidth]{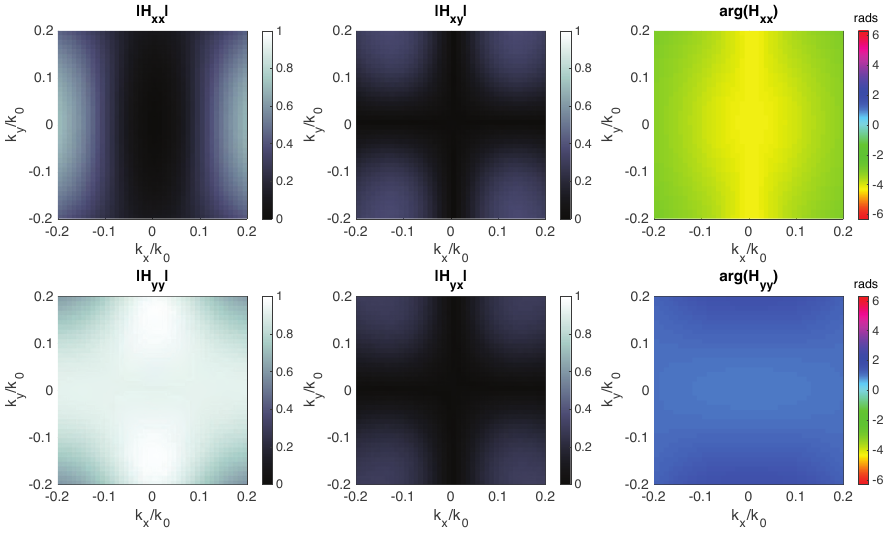}
    \caption{\textbf{2D optical transfer function of the metasurface}. The magnitude (first two columns) and phase (last column) of the optical transfer function tensor components at the operating wavelength.}
    \label{fig:2d_otf}
\end{figure*}

\begin{figure*}[!t]
   \centering
    \includegraphics[width=0.9\linewidth]{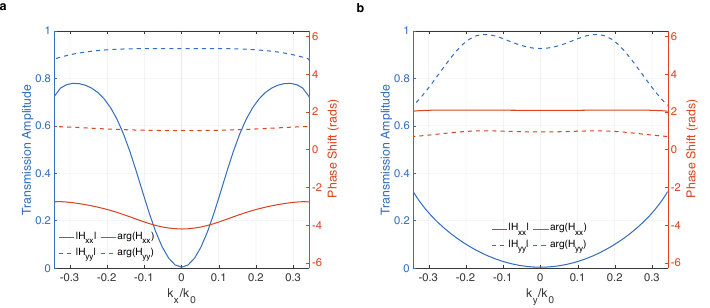}
    \caption{\textbf{1D optical transfer function of the metasurface}. (a) The magnitude and phase of the co-polarized components of the optical transfer function along $k_y = 0$. (b) Same as a but along $k_x = 0$.}
    \label{fig:1d_otf}
\end{figure*}

The magnitude and phase of the optical transfer function are given in Fig. \ref{fig:2d_otf}. It was computed as the angle-dependent transmission as a function of the polar and azimuthal angles of incidence ($\theta, \phi)$. The 1D slices through $k_x = 0$ and $k_y = 0$ are given in Fig. \ref{fig:1d_otf}. The results show that the capability of the metasurface to process input images along $k_y$ is weaker than that for $k_x$ under $x$-polarization. Moreover, the cross-polarized components of the optical transfer function are consistent with Ref. \cite{Shi2020SI}. They are non-zero along the diagonals in Fig. \ref{fig:2d_otf}(middle column), however the contribution to the output was measured to be small in the experiments. 

\subsection{Performance metrics}

The performance of the metasurface was quantified through various numerical metrics. These included the transmittance, reflectance, absorptance, quality factor, electric field enhancement, numerical aperture, angular response, polarization conversion efficiency, processing efficiency and the switching contrast. The results are tabulated in Table \ref{tbl:performance_sim} and can be compared to the experimental results in Table \ref{tbl:performance_exp}. The results show that the metasurface strongly reflects light when on-resonance with little absorption for normally incident, $x$-polarized illumination. The electric dipole resonance had a quality factor of 17.3, which produced an optical transfer function with a numerical aperture of approximately 0.17.

\begin{table}[!t]
  \caption{\textbf{Simulated performance metrics of the metasurface.} The performance metrics were obtained through simulations in COMSOL Multiphysics 6.1 at the operating wavelength of \SI{969}{\nano\meter}, corresponding to the electric dipole resonance of the metasurface. Here, $\Vec{E}_0$ denotes the electric field of the normally-incident, $x$-polarized illumination. The values $k_x^{\text{max}}$ and $k_x^{\text{min}}$ denote the spatial frequencies at the edges of the contrast zone. $E_t$ and $E_r$ denote the electric field transmitted through, and reflected from, the metasurface, respectively. $P_\text{cross}$ and $P_{\text{in}}$ denote the cross-polarized and input power, respectively. Lastly, $\Delta \lambda$ is the full width at half-maximum of the transmission minimum at the operating wavelength.}
  \label{tbl:performance_sim}
  \begin{tabular}{cccc}
    \hline
    Quantity & Label & Formula & Value  \\
    \hline
    Transmittance & $T$ & $|E_t/E_0|^2$ & 0.001 \\
    Reflectance & $R$ & $|E_r/E_0|^2$ & 0.993 \\
    Absorptance & $A$ & $1 - T - R$ & $\leq 0.006$ \\
    Quality factor & $Q$ & $\lambda / \Delta \lambda $ & 17.3 \\
    Field enhancement & $\Delta E$ & $\max(|E/E_0|) $ & 6.1 \\
    Numerical aperture & $\text{NA}$ & $(k_x^{\text{max}} - k_x^{\text{min}})/k_0$ & $\leq 0.17$ \\
    Angular response & $\Theta$ & $\Delta \lambda/ \lambda$ & 0.06 \\
    Polarization conversion efficiency & $\epsilon_{\text{pol}}$ & $100\% P_{\text{cross}}/P_{\text{in}}$ & 0.001\% \\
    Processing efficiency & $\eta$ & $|H_{xx}(k_x^{\text{max}},0)|$ & 79\% \\
    Switching contrast (amplitude) & $\delta$ & $H_{yy}(0,0) - H_{xx}(0,0)$ & 93\% \\
    Switching contrast (transmittance) & $\Delta$ & $T_y - T_x$ & 86\% \\
    \hline
  \end{tabular}
\end{table}


\subsection{Isolated resonator}

The optical response of an isolated silicon resonator of size given in the main article was investigated by implementing the formalism explained in section S1.2. The simulations were performed using the finite element method within the wave optics module of COMSOL Multiphysics 6.1. The scattering formulation was used to project the far-zone scattered electromagnetic field onto the vector spherical harmonics on a virtual sphere of radius $r_0 = \SI{750}{\nano\meter}$ surrounding the resonator. Sommerfeld boundary conditions and a perfectly matched layer were applied onto the virtual sphere to remove unwanted reflections into the system and to simulate an open domain. The finite meshing elements in the model had a a minimum size of \SI{5}{\nano\meter} and a maximum size of $\lambda/$\SI{8}{\nano\meter}. The perfectly matched layer had a series of 5 meshing elements. The scattering cross section was obtained by integrating the projections of the electromagnetic fields onto the vector spherical harmonics and using Eqs. (S10)-(S15).
                                                          
\begin{figure*}[!t]
   \centering
    \includegraphics[width=0.9\linewidth]{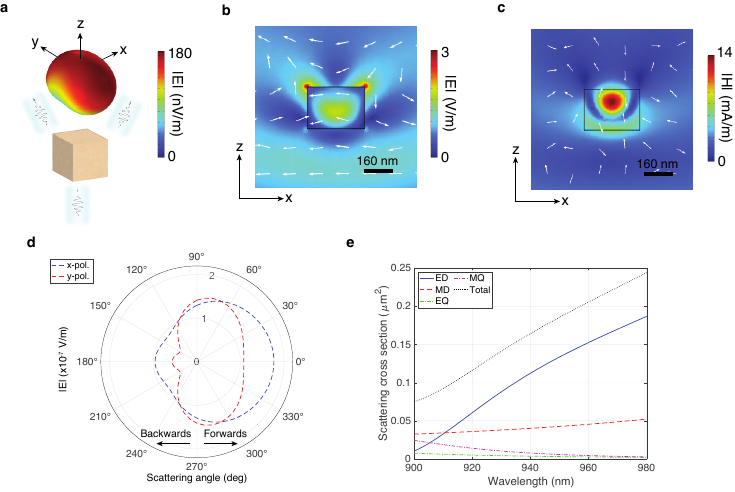}
    \caption{\textbf{Optical response of an isolated silicon resonator in air}. (a) The 3D radiation pattern. (b) The electric field strength (heat map) and direction (arrows) when on-resonance in the $xz$-plane. (c) Same as b but for the magnetic field. (d) The polar radiation plot in the $xz$-plane for $x$- and $y$-polarization. (e) A multipole decomposition of the scattering cross section as a function of the wavelength. The results in (a)-(c) and (e) were obtained with $x$-polarized illumination at the operating wavelength.}
    \label{fig:isolated_resonator}
\end{figure*}

The results are given in Fig. \ref{fig:isolated_resonator} for the case of normally incident, $x$-polarized plane waves illuminating the resonator at the operating wavelength. The 3D (Fig. \ref{fig:isolated_resonator}a) and polar radiation patterns (Fig. \ref{fig:isolated_resonator}d) show that majority of the light is concentrated in the forward direction. The electromagnetic field behavior in the $xz$-plane shows field enhancement within the resonator (Fig. \ref{fig:isolated_resonator}b-c). The multipole decomposition of the scattering cross section indicates that this is due to an electric dipole resonance (Fig. \ref{fig:isolated_resonator}e). 


\subsection{Imaging simulations}

\begin{figure*}[!t]
   \centering
    \includegraphics[width=0.9\linewidth]{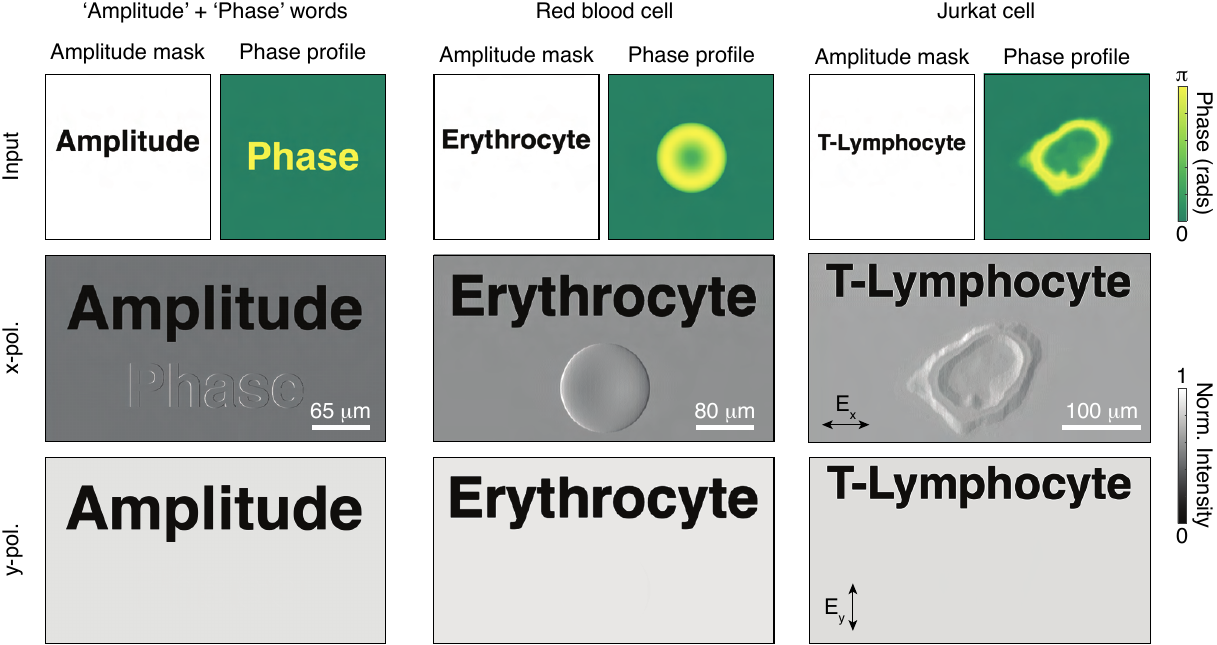}
    \caption{\textbf{Simulated tunable multi-modal microscopy with the metasurface}. (First row) The amplitude and phase masks of the input imaging target. (Second row) The output images obtained with $x$-polarized illumination. (Final row) The output images obtained with $y$-polarized illumination.}
    \label{fig:imaging_simulation}
\end{figure*}

Simulations of tunable multi-modal microscopy were performed in MATLAB R2022a by the formulations presented in the `Tunable multi-modal microscopy - principles' section of the main article and in the section S1.1. The optical transfer function of the metasurface that was calculated in COMSOL Multiphysics 6.1 (Fig. \ref{fig:2d_otf}) was used to model the optical response of the metasurface in Eq. \eqref{eq:otf-conv-thm}. Three imaging targets were used in the simulations and experiments. The first included an opaque amplitude mask with the word `Amplitude' and a phase profile of the word `Phase' (Fig. \ref{fig:imaging_simulation} - top left), which was presented in the main article. The second included the opaque word `Erythrocyte' and the phase profile of a human red blood cell (Fig. \ref{fig:imaging_simulation} - top middle). Some of the experimental imaging results are presented in Fig. 1a of the main article. The phase profile of the red blood cell was modelled using the empirical optical properties provided in Ref. \cite{Evans1972SI}. Lastly, the third imaging target included the opaque word `T-Lymphocyte' and the phase profile of a leukaemic Jurkat cell (Fig. \ref{fig:imaging_simulation} - top right) \cite{Schneider1977SI}. The phase profile of the Jurkat cell was modelled using the empirical optical properties from Ref. \cite{Zhang2017SI} and the imaging results for both simulation and experiment are presented here. 

The simulated imaging results are presented in Fig. \ref{fig:imaging_simulation}. The images obtained with $x$-polarized illumination contained both of the transparent and opaque features of the input imaging target (Fig. \ref{fig:imaging_simulation} - middle row). Each of the phase profiles have a pseudo-3D appearance in the images. On the other hand, only the opaque words are visible in the images obtained with $y$-polarized illumination (Fig. \ref{fig:imaging_simulation} - bottom row).

\section{Optical configurations}


\subsection{Polarization}

\begin{figure*}[!t]
   \centering
    \includegraphics[width=0.7\linewidth]{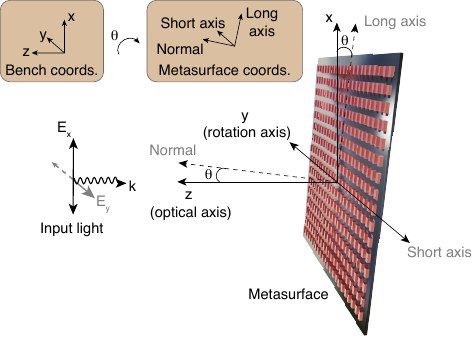}
    \caption{\textbf{Polarization}. Schematic of how the polarization is defined and referred to in this article.}
    \label{fig:polarization}
\end{figure*}

The metasurface operates under the excitation by linearly polarized illumination. As in the main article, we define two coordinate systems. The first is the global coordinate system with respect to the optical bench in the experiment. The second is with respect to the metasurface defined by the long axis of the silicon blocks, the short axis and the normal to the metasurface. Assume that the illumination propagates along the global $z$-axis (optical axis). The metasurface was placed in the global $xy$-plane with its long axis aligned with the global $x$-axis. It was then tilted by \SI{2}{\degree} about the global $y$- (short) axis. Therefore, with this configuration, $x$-polarized illumination leads to the production of phase contrast images, while $y$-polarized illumination leads to bright field images. A schematic of the polarization is given in Fig. \ref{fig:polarization}. 

In the imaging experiments, the global $x$-axis was along the vertical direction and the global $y$-axis was along the horizontal direction. The metasurface was oriented with its long axis along the vertical direction (at normal incidence), as depicted in Fig. \ref{fig:polarization}. To change the polar angle of incidence $\theta$, the metasurface was tilted about the horizontal direction. The illumination was $x$-polarized when the input polarizer was along the vertical direction. On the other hand, the illumination was $y$-polarized when the input polarizer was horizontal. The effects of de-polarization and/or cross-polarization by the spatial light modulator (SLM), imaging samples, metasurface and optical components were measured to be negligible. For example, the slight deviations on the polarization set by the polarizer due to diffraction from the SLM was a weak effect. 


\subsection{Spectrometry}

\begin{figure*}[!t]
   \centering
    \includegraphics[width=0.9\linewidth]{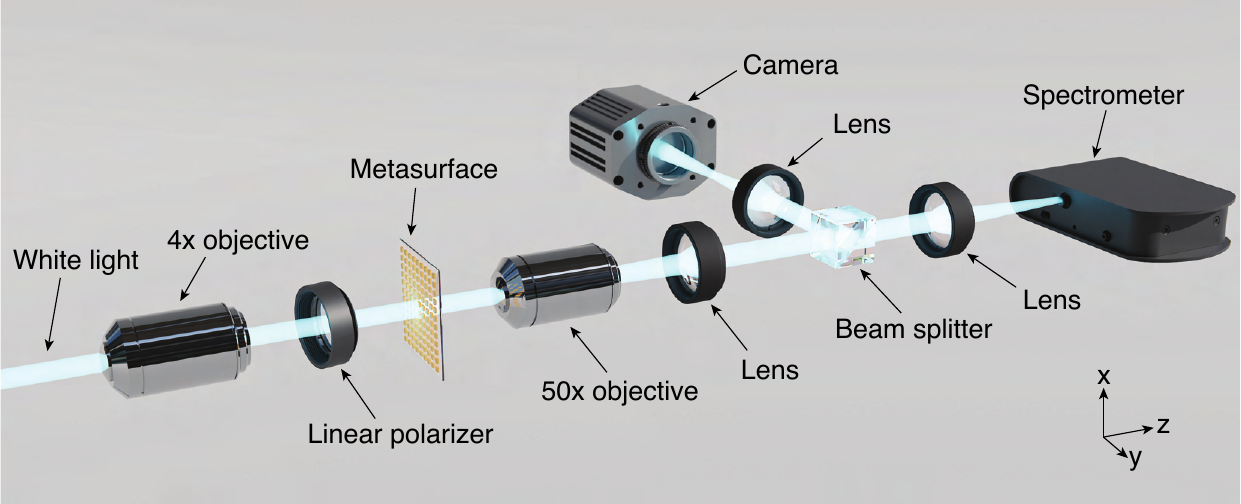}
    \caption{\textbf{Normal-incidence spectroscopy configuration}. White light from a halogen lamp was collimated by a microscope objective and linearly polarized. The light transmitted by the metasurface was collected by a second microscope objective and a lens. A second lens focused the transmitted light onto a spectrometer. A beam-splitter was used to simultaneously relay an image of the metasurface onto a camera by a lens.}
    \label{fig:spectroscopy_exp}
\end{figure*}

\begin{figure*}[!t]
   \centering
    \includegraphics[width=0.9\linewidth]{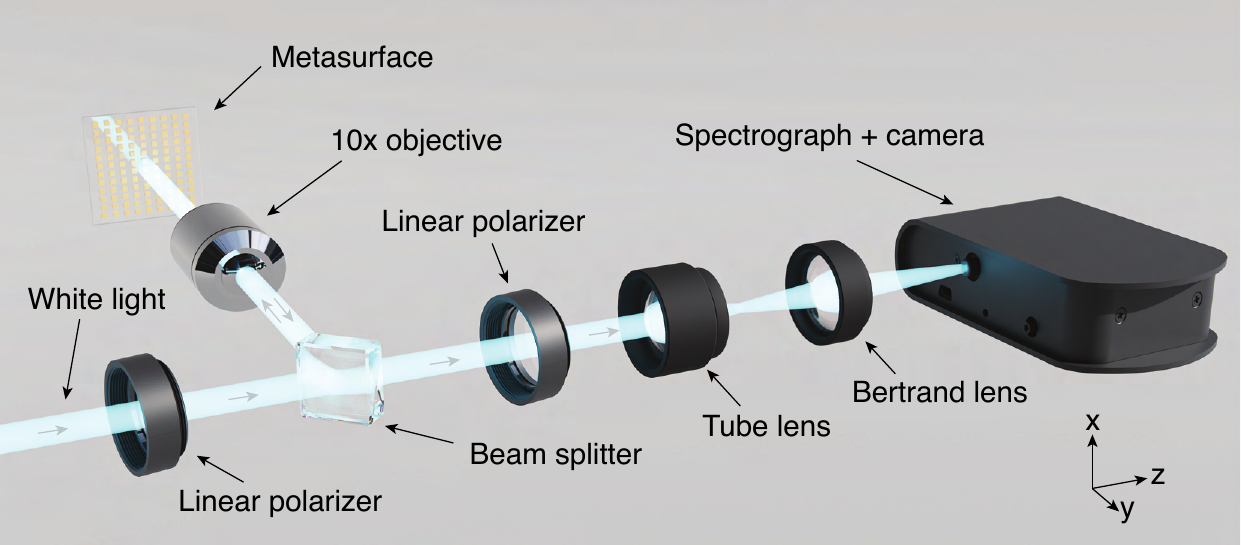}
    \caption{\textbf{Angle-dependent spectroscopy via back focal plane imaging}. White light from a halogen lamp was linearly polarized and illuminated the metasurface through a microscope objective. The reflected light from the objective back focal plane was imaged onto a spectrograph/CCD couple with a tube lens and Bertrand lens pair. An analyzer was used before the spectrograph to select specific components of the optical transfer function tensor.}
    \label{fig:bfp_exp}
\end{figure*}

The normal-incidence and angle-dependent spectroscopy experiments are detailed in the Methods section of the main article. Here, schematics of their experimental configurations are shown in Figs. \ref{fig:spectroscopy_exp} and \ref{fig:bfp_exp}.

\subsection{Imaging with a spatial light modulator}

\begin{figure*}[!t]
   \centering
    \includegraphics[width=0.9\linewidth]{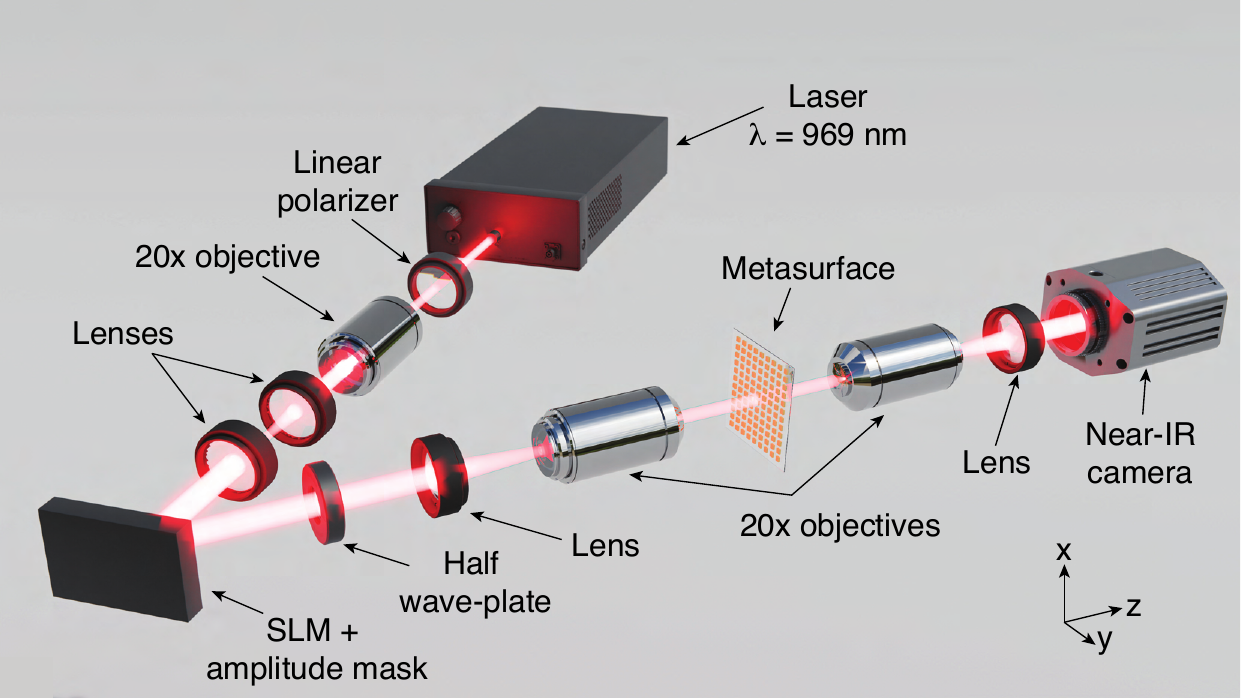}
    \caption{\textbf{Experimental configuration for imaging involving a spatial light modulator}. The light from a super-continuum source was linearly polarized and collimated by a microscope objective and two lenses. The lenses expanded the beam to fill the aperture window of the SLM. An amplitude mask was placed in front of the SLM. The light reflected by the SLM was focused into a microscope objective by a lens. The microscope objective produced a small collimated beam containing a de-magnified image of the SLM aperture. The metasurface was placed in the image plane. A half wave-plate was placed before the lens to change the polarization. A microscope objective and lens relayed a magnified image onto a camera.}
    \label{fig:slm_exp}
\end{figure*}

The experimental imaging results involving the SLM presented were obtained with the configuration illustrated in Fig. \ref{fig:slm_exp}. This schematic is also shown in Fig. 3 of the main article.


\subsection{Biological microscopy}

\begin{figure*}[!t]
   \centering
    \includegraphics[width=0.9\linewidth]{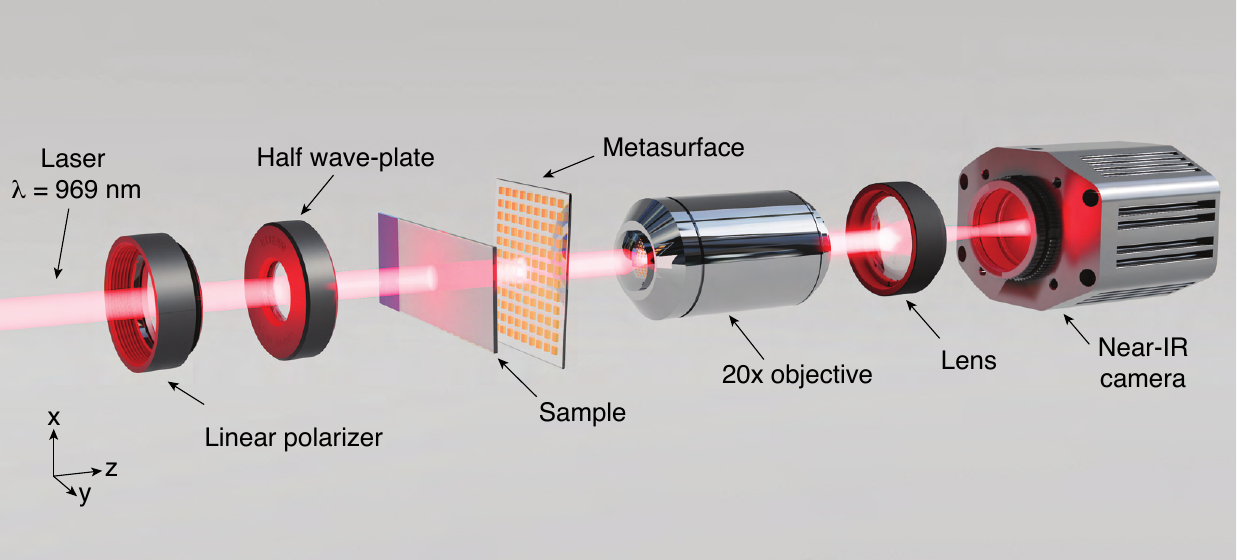}
    \caption{\textbf{Bio-imaging configuration}. The light from a super-continuum source was linearly polarized and a half wave-plate was used to control the polarization. The light illuminated a sample and the metasurface was placed after the sample. A microscope objective and lens relayed a magnified image onto a camera.}
    \label{fig:bio_exp}
\end{figure*}

The experimental biological imaging results with the metasurface presented in the main article were obtained with the configuration illustrated in Fig. \ref{fig:slm_exp}. This schematic is also shown in Fig. 4 of the main article.
 


\section{Additional experimental data}


\subsection{Metasurface fabrication}

\begin{figure*}[!t]
   \centering
    \includegraphics[width=0.9\linewidth]{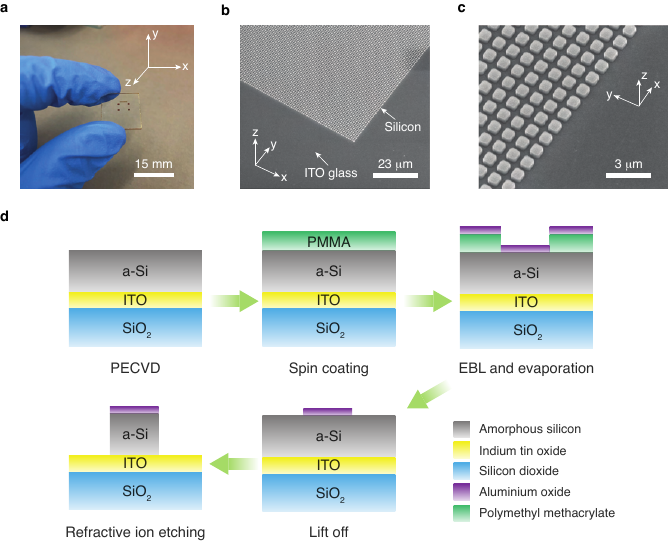}
    \caption{\textbf{Fabrication of the metasurface}. (a) A photo of the metasurface fabricated onto a coverslip. (b),(c) Scanning electron microscope images of the metasurface. (d) A flowchart of the fabrication process. Plasma-enhanced chemical vapor deposition is abbreviated to PECVD.}
    \label{fig:metasurface-SI}
\end{figure*}

A flowchart of the fabrication process and pictures of the metasurface are given in Fig. \ref{fig:metasurface-SI}. The metasurface was fabricated on a \SI{10}{\milli\meter}-thick glass substrate that was \SI{15}{\milli\meter} by \SI{15}{\milli\meter} in size. A thin \SI{23}{\nano\meter} layer of indium tin oxide (ITO) was coated onto the glass substrate to be used as a conducting protector for the electron beam lithography process. Similarly, a thin \SI{10}{\nano\meter} layer of aluminium oxide (Al$_2$O$_3$) was coated onto the silicon to be used as a hard protective mask for the etching process.  


\subsection{Optical characterization}

\begin{figure*}[!t]
   \centering
    \includegraphics[width=0.9\linewidth]{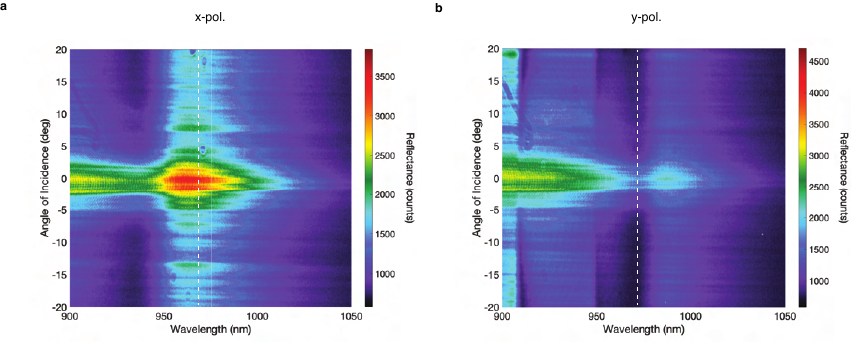}
    \caption{\textbf{Angle-dependent spectroscopy via back focal plane imaging}. (a) 2D heat map of the reflectance counts from the metasurface as a function of the wavelength and angle of incidence for $px$-polarization. (b) Same as a but for $y$-polarization. The white dashed lines indicate the operating wavelength. }
    \label{fig:bfp_results}
\end{figure*}

The angle-dependent spectroscopy experiments were performed with the back focal plane imaging configuration depicted in Fig. \ref{fig:bfp_exp} and detailed in the Methods section of the main article. The metasurface was oriented such that the angular range would be along the plane of incidence defined by the long axis of the metasurface. The results (Fig. \ref{fig:bfp_results}) show the reflectance counts as a function of the wavelength and angle for input $x$- and $y$-polarized illumination. The former shows strong reflection at normal incidence at the operating wavelength, which decreases in strength as the angle increases. Meanwhile, the latter has a relatively low reflection. Plots of the line profiles along the line segments in Fig. \ref{fig:bfp_results} are given in Fig. 2b-ii of the main article in the form of transmission amplitude data.


\subsection{Performance metrics}

\begin{table}[!t]
  \caption{\textbf{Experimental performance metrics of the metasurface.} The experimental performance metrics obtained at the operating wavelength of \SI{969}{\nano\meter}.}
  \label{tbl:performance_exp}
  \begin{tabular}{cccc}
    \hline
    Quantity & Label & Formula & Value  \\
    \hline
    Transmittance & $T$ & $|E_t/E_0|^2$ & $\geq 0.096$ \\
    Reflectance & $R$ & $|E_r/E_0|^2$ & $\leq 0.931$ \\
    Quality factor & $Q$ & $\lambda / \Delta \lambda $ & 19.29 \\
    Numerical aperture & $\text{NA}$ & $(k_x^{\text{max}} - k_x^{\text{min}})/k_0$ & $\leq 0.17$ \\
    Angular response & $\Theta$ & $\Delta \lambda/ \lambda$ & 0.05 \\
    Processing efficiency & $\eta$ & $|H_{xx}(k_x^{\text{max}},0)|$ & 78\% \\
    Switching contrast (amplitude) & $\delta$ & $H_{yy}(0,0) - H_{xx}(0,0)$ & 60\% \\
    Switching contrast (transmittance) & $\Delta$ & $T_y - T_x$ & 71\% \\
    \hline
  \end{tabular}
\end{table}

The experimental performance of the metasurface was quantified through the metrics presented in Table \ref{tbl:performance_exp}. These included the transmittance, reflectance, quality factor, numerical aperture, angular response, processing efficiency and the switching contrast. The results are tabulated in Table \ref{tbl:performance_exp} and can be compared to the simulations in Table \ref{tbl:performance_sim}.


\subsection{Imaging experiments}

\begin{figure*}[!t]
   \centering
    \includegraphics[width=0.9\linewidth]{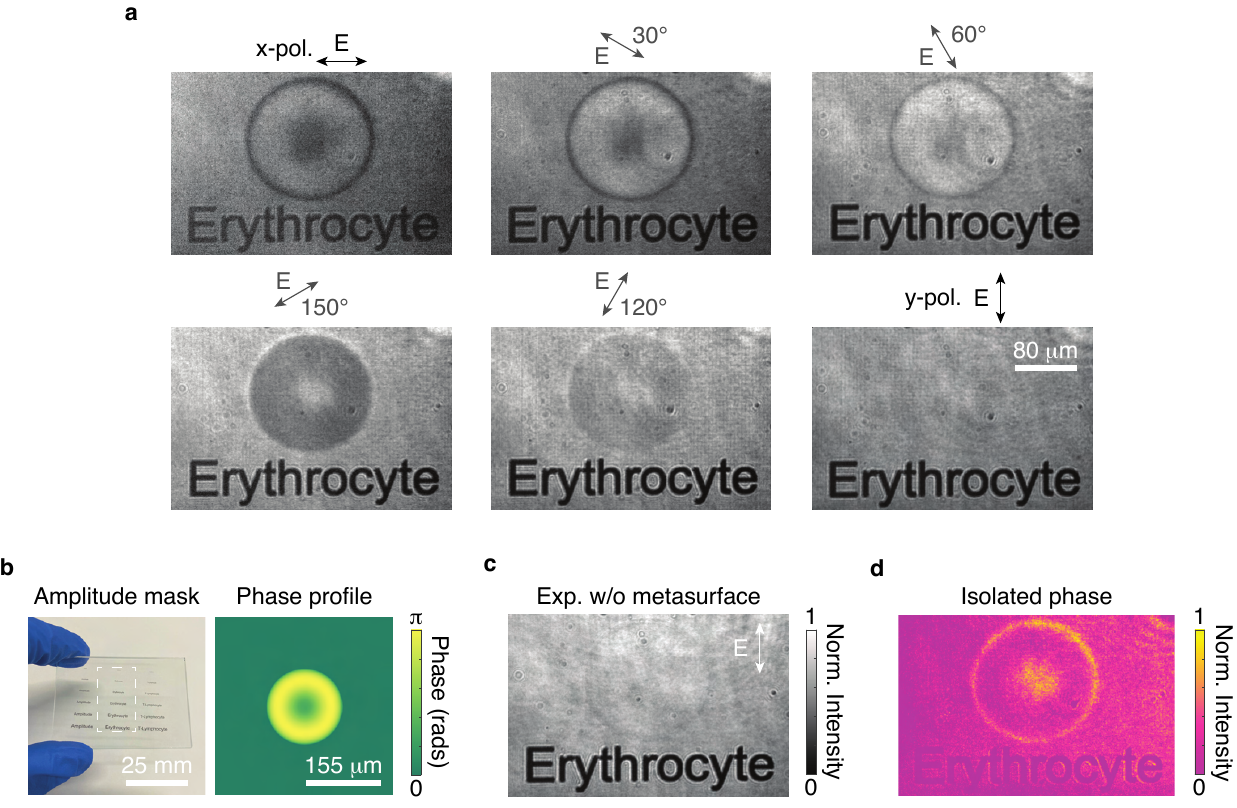}
    \caption{\textbf{Tunable multi-modal microscopy experiments of samples produced by a spatial light modulator - red blood cell}. (a) The experimental images obtained with the metasurface for different polarization states of the illumination at the operating wavelength. (b) The input imaging target included an amplitude mask and a phase profile on the spatial light modulator. (c) The experimental image obtained in the absence of the metasurface. (d) The image obtained by subtracting the image obtained with $x$-polarized illumination from the image obtained with $y$-polarized illumination.}
    \label{fig:SLM_rbc}
\end{figure*}

\begin{figure*}[!t]
   \centering
    \includegraphics[width=0.9\linewidth]{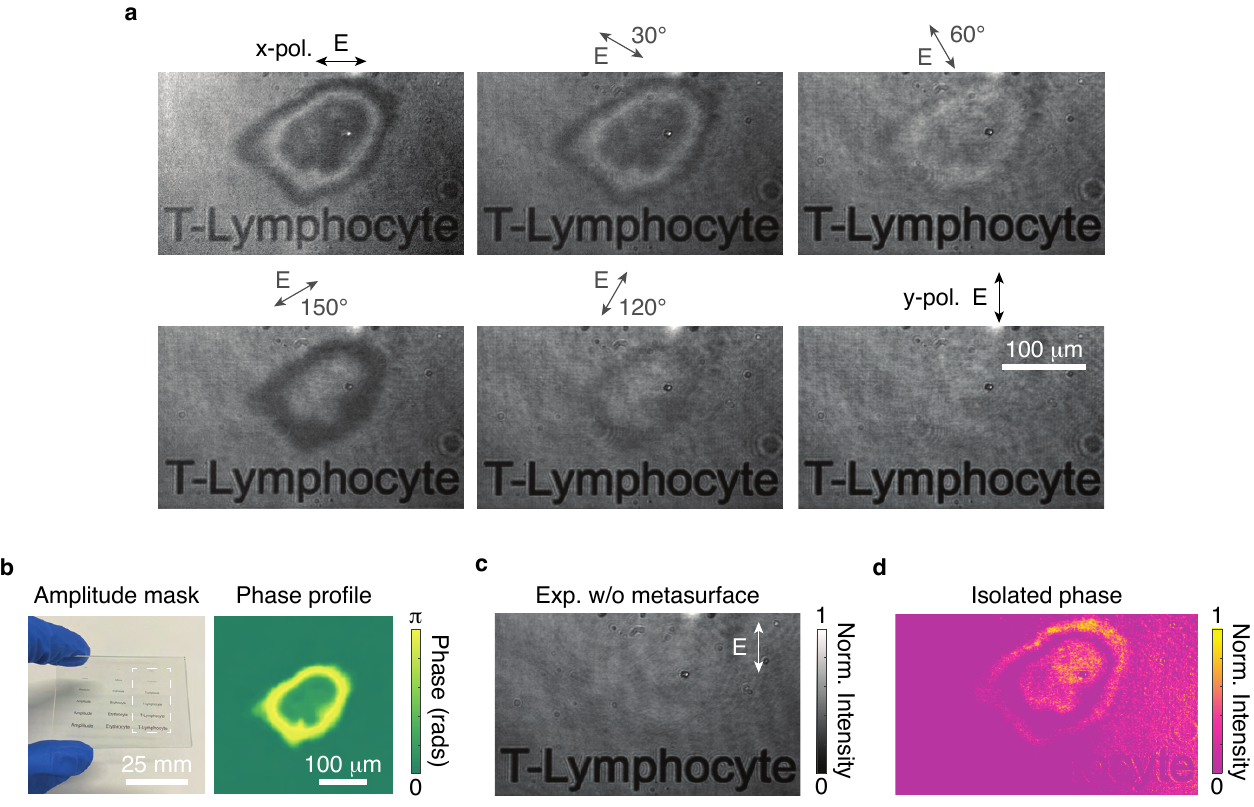}
    \caption{\textbf{Tunable multi-modal microscopy experiments of samples produced by a spatial light modulator - Jurkat cell}. (a) The experimental images obtained with the metasurface for different polarization states of the illumination at the operating wavelength. (b) The input imaging target included an amplitude mask and a phase profile on the spatial light modulator. (c) The experimental image obtained in the absence of the metasurface. (d) The image obtained by subtracting the image obtained with $x$-polarized illumination from the image obtained with $y$-polarized illumination.}
    \label{fig:SLM_jurkat}
\end{figure*}

The multi-modal microscopy experiments involving the SLM were performed with the three different imaging targets set out in section S2.4. The imaging results with the `Amplitude' and `Phase' words in the imaging target were presented in Fig. 3 of the main article. The imaging results of the red blood cell are given in Fig. \ref{fig:SLM_rbc} and the results of the Jurkat cell are given in Fig. \ref{fig:SLM_jurkat}.

The red blood cell (Fig. \ref{fig:SLM_rbc}b) and Jurkat cell (Fig. \ref{fig:SLM_jurkat}b) were visualized in the images obtained under $x$-polarized illumination (Figs. \ref{fig:SLM_rbc}a and \ref{fig:SLM_jurkat}a). The contrast of these features had a pseudo-3D appearance. On the other hand, the words `Erythrocyte' (Fig. \ref{fig:SLM_rbc}b) and `T-Lymphocyte' (Fig. \ref{fig:SLM_jurkat}b) were visible in all of the images obtained in the experiment (Figs. \ref{fig:SLM_rbc}a and \ref{fig:SLM_jurkat}a), independent of the polarization. In particular, only these features were visible in the images obtained under $y$-polarization. These images were similar to the images taken in the absence of the metasurface (Figs. \ref{fig:SLM_rbc}c and \ref{fig:SLM_jurkat}c). The images were combined to qualitatively isolate the phase by subtracting the images obtained with $x$-polarization from those obtained with $y$-polarization (Figs. \ref{fig:SLM_rbc}d and \ref{fig:SLM_jurkat}d).


\subsection{Biological imaging}

As discussed in the main article, the thickness of the breast tissue sample was a challenge for the imaging modalities used in the experiments. The metasurface and the DIC microscope were capable of highlighting the architectural structure within the tissue, even at the higher thickness provided by the microtomy blade sectioning of the whole tissue blocks. This capacity can allow for the identification of normal architectural patterns within the tissue and those which are abnormal, e.g. tumours often form aberrant architectural structural arrangements. However, identifying more complex structures or lower order cellular detail is uncertain, as the imaged area included only a small number of structures in a simple and benign arrangement. 

Instead, a more ideal application would be the preparation of cellular material rather than formalin-fixed, paraffin embedded in-situ tissue that requires a thicker tissue sample. In this case, the imaging samples would be generated by smearing fresh tissue to directly transfer individual and clusters of cells onto a glass slide, rather than an entire section of tissue. This produces a thinner sample with cells juxtaposed on a background slide, rather than remaining embedded in their native tissue environment. These thinner samples would provide a greater differential between individual and groups of cells and the background glass slide. Furthermore, it may facilitate a better generation of cellular level imaging data.  


\subsection{Experimental videos}

Two experimental videos are included as supplementary material. The first (Video S1) is a recording of an experiment made on the polarization switching ability of the metasurface. By using the experimental configuration detailed in the `Multi-modal microscopy - biological samples and polarization switching' sub-section of the Methods section of the main article, an image of the metasurface was captured by a camera under normally incident illumination at the operating wavelength. The video shows how the performance of the metasurface changes as a function of the polarization. 

The second (Video S2) video is a recording of an experiment made for imaging \textit{C. elegans} with the metasurface. The video shows a live nematode moving in the sample, demonstrating the capacity for the metasurface to be used for dynamic monitoring of samples.



\end{document}